\documentclass[conference]{IEEEtran}
\usepackage{graphicx,epsfig,amsmath,amssymb,bm}
\usepackage{amssymb,balance}
\usepackage{amsmath}
\usepackage{amsthm} 
\usepackage{tikz}
\usepackage{amssymb}
\usepackage{algorithmicx}
\usepackage{algpseudocode}
\newtheorem{defn}{Definition}[section]
\newtheorem{theorem}{Theorem}
\def\LSB{\left[}        
\def\RSB{\right]}       
\def\LB{\left(}         
\def\RB{\right)}        

\newfont{\bbb}{msbm10 scaled 500}

\newfont{\bb}{msbm10 scaled 1100}
\newcommand{\CC}{\mbox{\bb C}}
\newcommand{\RR}{\mbox{\bb R}}

\newcommand{\av}{{\bf a}}

\newcommand{\cv}{{\bf c}}

\newcommand{\fv}{{\bf f}}
\newcommand{\gv}{{\bf g}}

\newcommand{\lv}{{\bf l}}

\newcommand{\nv}{{\bf n}}

\newcommand{\pv}{{\bf p}}

\newcommand{\rv}{{\bf r}}

\newcommand{\uv}{{\bf u}}

\newcommand{\vv}{{\bf v}}
\newcommand{\xv}{{\bf x}}

\newcommand{\zv}{{\bf z}}

\newcommand{\Am}{{\bf A}}
\newcommand{\Bm}{{\bf B}}

\newcommand{\Dm}{{\bf D}}

\newcommand{\Hm}{{\bf H}}
\newcommand{\Id}{{\bf I}}

\newcommand{\Wm}{{\bf W}}

\newcommand{\thetav}{\hbox{\boldmath$\theta$}}

\newcommand{\Gammam}{\hbox{\boldmath$\Gamma$}}

\newcommand{\rank}{{\hbox{rank}}}

\newcommand{\defines}{{\,\,\stackrel{\scriptscriptstyle \bigtriangleup}{=}\,\,}}

\usepackage{bbm}
\usepackage{subcaption}
\usepackage{tabularx}
\usepackage{multirow}
\usepackage{verbatim}

\def\argmax{\operatornamewithlimits{arg\,max}}

\newtheorem{proposition}{Proposition}

\newcommand{\beqa}{\begin{eqnarray}}
\newcommand{\eeqa}{\end{eqnarray}}
\newcommand{\dsp}{\displaystyle}

\ifCLASSINFOpdf

\else

\fi

\begin{document}

\title{Cost-Benefit Analysis of Moving-Target Defense in Power Grids}
\IEEEoverridecommandlockouts 
\author{
\IEEEauthorblockN{Subhash Lakshminarayana\IEEEauthorrefmark{1} and David K.Y. Yau \IEEEauthorrefmark{1}\IEEEauthorrefmark{2}} 
\IEEEauthorblockA{\IEEEauthorrefmark{1}
Advanced Digital Sciences Center, Illinois at Singapore, Singapore 138602 \\
\IEEEauthorrefmark{2}
Singapore University of Technology and Design, Singapore 487372\\ 
Email: \IEEEauthorrefmark{1}subhash.l@adsc-create.edu.sg, \IEEEauthorrefmark{2} david\_yau@sutd.edu.sg}

\thanks{
This work was supported by the National Research Foundation (NRF), Prime Minister's Office, Singapore, under its National Cybersecurity R\&D Programme (Award No. NRF2014NCR-NCR001-31) and administered by the National Cybersecurity R\&D Directorate.}
}

\maketitle

\begin{abstract}
We study moving-target defense (MTD) that actively perturbs transmission line reactances to thwart stealthy false data injection (FDI) attacks against state estimation in a power grid. Prior work on this topic has proposed MTD based on randomly selected reactance perturbations, but these perturbations cannot guarantee effective attack detection. To address the issue, we present formal design criteria to select MTD reactance perturbations that are truly effective. However, based on a key optimal power flow (OPF) formulation, we find that the effective MTD may incur a non-trivial operational cost that has not hitherto received attention. Accordingly, we characterize important tradeoffs between the MTD's detection capability and its associated required cost. Extensive simulations, using the MATPOWER simulator and benchmark IEEE bus systems, verify and illustrate the proposed design approach that for the first time addresses both key aspects of cost and effectiveness of the MTD.
\end{abstract}

\IEEEpeerreviewmaketitle

\section{Introduction}
Cyber attacks against critical infrastructures can lead to severe disruptions. The December 2015 attack against the Ukraine's power grid was a real-world example, which caused power outages for a large number of customers for hours \cite{Ukraine2016}. These attacks were typically crafted by sophisticated attackers, sometimes with national backing,
who managed to spend considerable time inside a system to learn its operational details, and accordingly designed the injection of malicious data/control to disrupt its operations \cite{Ukraine2016:Analysis}. It is thus imperative to design counteracting defense approaches to defeat the knowledgeable attackers. Moving-target defense (MTD)~\cite{MTD-DHS} is a defense approach that has received increasing attention. It is based on dynamically changing the system parameters that attackers need to target for customizing their attacks, in order to invalidate the attackers' prior knowledge of the system and render ineffective any of their prior designed strategies. It has the potential to make it extremely difficult or impossible for would-be attackers to keep up with the system dynamics.

In this paper, we focus on \emph{false data injection} (FDI) attacks against state estimation (SE) in power grids. SE is a key method for grid operators to obtain a best estimate of the system state from noisy sensor measurements collected via a supervisory control and data acquisition (SCADA) system, for example. Its output is used in critical applications such as economic dispatch (for profits) and contingency analysis (for reliability). A bad data detector (BDD) associated with the SE is often deployed for identifying bad data (e.g., sensor anomalies and FDI attacks) to ensure trustworthy results. However, it has been shown~\cite{Liu2009} that FDI attacks crafted using detailed knowledge of a power grid's topology and the reactance settings of its transmission lines can bypass the BDD and remain stealthy. Such an undetected attack can have severe consequences, e.g., trips of transmission line breakers or unsafe frequency excursions \cite{RenLRAttacksTPDS2012,RuiAGC}.

To strengthen the BDD, it has been shown that if a carefully chosen subset of the sensors can be well protected (e.g., by  tamper-proof and encryption-enabled PLCs), or if a key subset of the state variables can be independently and reliably verified by phasor measurement units (PMUs) deployed at strategically chosen locations, then a BDD-bypassing FDI attack becomes impossible~\cite{Bobba2010,Dan2010,KimPoorProtection2011}. However, a major revamp of the basic sensing infrastructure can be quite expensive (e.g., PMU has high cost~\cite{PMUCost}) or infeasible for the many existing legacy systems whose life cycles often last decades and which are not expected to retire for the foreseeable future. Alternatively, FDI attacks can be significantly mitigated by MTD that invalidates the knowledge attackers used for crafting their prior attacks, specifically by active perturbation of the grid's transmission line reactance settings in our application context~\cite{Morrow2012,Davis2012,RahmanMTD2014}. This approach is practical because of current D-FACTS devices capable of active impedance injection~\cite{DFACTS2007}. Because of their low cost and ease and flexibility of deployment, they are being increasingly installed in existing alternating-current (ac) transmission networks to control power flows \cite{RogersDFACTS2008}. 

Prior work on MTD for FDI attacks against power grid SE has two important limitations, which are related. First, the MTD is implemented by selecting a random subset of transmission lines and introducing similarly random perturbations to their reactance settings~\cite{RahmanMTD2014}. There are no known conditions for the MTD perturbations to be truly effective. An important finding of this paper is that the randomly selected perturbations do not necessarily guarantee effective detection. Rather, a perturbation must satisfy certain design criteria that we will make clear (in Section~\ref{sec:MTD_Att_Det}), or FDI attacks crafted using (outdated) system knowledge before the perturbation will remain stealthy after it. Second, without an adequate characterization of effective MTD, prior work has not been able to address explicitly the associated cost involved. Rather, it is assumed that the MTD can be always constrained to have negligible or some ``low enough'' operational cost~\cite{RahmanMTD2014,Morrow2012}. However, MTD designed with any absolute cost constraints will not be useful if the MTD does not perform. It is thus critical to understand the inherent cost-benefit tradeoff of the MTD to  accordingly inform system operators (SOs) in their choice of security policies, which is a key objective of this paper.

To achieve our goal, we analyze the problem of selecting MTD reactance perturbations that jointly consider their effectiveness (i.e., capability of attack detection) and operational cost (i.e., economic inefficiency). As in prior work, we assume that the attacker has learned the system configuration initially and uses this knowledge to craft stealthy FDI attack vectors, but the attacker cannot track the reactance perturbations 
without significant delays. In this setting, large MTD perturbations will cause the actual system to deviate significantly from the attacker's prior knowledge, so that a large majority of the previously stealthy FDI attacks will now likely become detectable. Conversely, however, the large perturbations will also cause the power grid to operate significantly away from the optimal state, thereby incurring a significantly higher economic cost. On the other hand, smaller perturbations will be less expensive, but risk more undetected attacks. The general cost-benefit tradeoff is thus interesting.

In this paper, we address the cost-benefit tradeoff of the MTD by formulating its perturbation selection as a constrained optimization problem, namely minimization of the operational cost subject to a given effectiveness constraint. The operational cost is quantified as the increment due to the MTD over the cost achieved at optimal power flow (OPF) of the system without MTD. This cost is always non-negative. The effectiveness is quantified as the fraction of prior stealthy FDI attacks (i.e., those before the MTD perturbation) that will become detectable by the BDD after the perturbation. It is difficult to give an exact analysis of the effectiveness. 
We will instead employ a heuristic metric that effectively invalidates the attacker's knowledge required to bypass the BDD. Extensive simulation results show that the heuristic metric effectively approximates the true metric.

We use a direct-current (dc) power flow model to approximate power flows in an alternating-current (ac) grid. This approach is widely adopted and well justified in power system research (e.g., \cite{Liu2009,Bobba2010,RahmanMTD2014}). Under the dc model, 
the OPF cost corresponds mainly to the cost of generation dispatch.
Moreover, the sensor measurements are linearly related to the system state through a \emph{measurement matrix}, which in turn depends on the power grid topology and the reactance of the transmission lines. Naturally, perturbing a branch reactance will alter the measurement matrix correspondingly. A key observation in our analysis is that the MTD's effectiveness and operational cost are   related to the separation between the column spaces of the measurement matrices before and after the MTD. While the effectiveness is enhanced by increasing the separation between the two column spaces, the operational cost increases. Therefore, different degrees of separation between the two spaces provide a spectrum of balance between the two metrics.

We note that, in light of our deliberate cost analysis of the MTD, the MTD can be viewed as a form of insurance against possible FDI attacks. Such insurance requires an ongoing payment of ``premiums'' irrespective of whether an attack occurs or not. However, in the event of an attack, which may be accumulatively extremely expensive if allowed to persist indefinitely because of lack of detection, the insurance can provide a much needed hedge against the damage. In actual deployments, whether to procure such insurance (i.e., turn on the MTD or not) is likely a matter of diverse factors such as institutional policies (including the institution's attitude towards risk taking), estimated vulnerability to attacks or likelihood of attacks, and the cost-benefit tradeoff specific to the power grid in question. This paper sheds light on tradeoffs in the key technical problem, which serves as an important reference basis for the other questions. Nevertheless, it does not attempt to answer all the questions, particularly policy questions, that are also interesting.

The main contributions of the paper are summarized as follows:
\begin{itemize}
\item We derive conditions for an MTD reactance perturbation to ensure that no FDI attacks crafted based on the outdated (pre-perturbation) system configuration will remain stealthy after the perturbation.  
\item When the reactance adjustment capability of D-FACTS is insufficient 
for achieving the above condition, we present heuristic design criteria for selecting MTD perturbations 
that can still highly likely achieve effective attack detection.
\item We characterize the tradeoff between the MTD's effectiveness and its operational cost in a constrained optimization framework. Additionally, we present extensive simulation results using the realistic MATPOWER simulator for benchmark IEEE bus systems to verify and illustrate the tradeoff.
\end{itemize}

The remainder of this paper is organized as follows. Section~\ref{sec:Related_Work} reviews related work. Section~\ref{sec:Prelim} introduces the preliminaries. Section~\ref{sec:MTD} explains the attacker and the defender model. Sections~\ref{sec:MTD_Att_Det} and \ref{sec:MTD_Cost} analyze the MTD's effectiveness and its cost-benefit tradeoff. Section~\ref{sec:Sim_Res} presents simulation results. Section~\ref{sec:Conc} concludes. The technical proofs can be found in Appendices~A,B and C.

\section{Prior Work}
\label{sec:Related_Work}
Recent work \cite{Liu2009} analyzed the condition for bypassing the BDD of SE and proposed a technique to construct BDD-bypassing FDI attacks using complete knowledge of the power grid topology and the branch reactances. Subsequent research  \cite{TeixeiraCDC2010} showed that such attacks can be constructed using partial knowledge of the power grid topology. 
However, the knowledge of power grid topology is difficult to obtain in practice. Recent work \cite{KimTong2015,PoorBlind2013} showed that BDD-bypassing attacks can also be crafted using the eavesdropped measurement data only. 
The impact of such stealthy FDI attacks on system efficiency and safety were investigated. 
In particular, the economic impact of FDI attacks were studied in \cite{Sinopoli_MarketOp2011} and \cite{RenLoadRedis2011}. 
Reference \cite{RuiAGC} showed that the attacker can drive the power system frequency to unsafe levels by injecting a sequence of carefully-crafted FDI attacks. 

To address BDD's vulnerability, defense mechanisms based on protecting a strategically-selected set of sensors and their data links were proposed \cite{Bobba2010,Dan2010,KimPoorProtection2011}. 
The use of generalized likelihood ratio test was proposed to detect FDI attacks when the adversary has access to only a few meters in \cite{Kosut2011}. Reference \cite{LiuSparse2014} presented a sparse optimization based approach to separate nominal power grid states and anomalies.

The concept of MTD was originally proposed for enterprise networks based on changing the IT features of devices such as end hosts' IP addresses and port numbers, the routing paths between nodes, etc. \cite{Antonatos:2007,Jafarian:2012}. More recent work 
has proposed MTD in power systems by changing its physical characteristics \cite{Morrow2012,Davis2012,RahmanMTD2014}. In particular, \emph{on-going} FDI attacks can be detected by introducing reactance perturbations that are known only to the defender (SO) \cite{Morrow2012}, since the change in sensor measurements (after the perturbations) will be different from its predicted value based on the power flow model (due to the attack). It has also been shown that stealthy FDI attacks can be precluded by actively perturbing the branch reactances to invalidate the attacker's knowledge  \cite{RahmanMTD2014}. We similarly consider MTD for power systems in this paper.
Compared with the prior work, ours is the first to jointly consider the MTD's effectiveness
and its operational cost. We provide hitherto unavailable formal design criteria for selecting effective MTD reactance perturbations, and expose important tradeoffs between the effectiveness and operational cost.

\section{Preliminaries}
\label{sec:Prelim}

\subsection*{Power Grid Model}
We consider a power network that is characterized by a set $\mathcal{N} = \{1,\dots,N\}$ of buses, 
$\mathcal{L} = \{1,\dots,L\}$ of transmission lines (an example of the 4~bus power system is shown in Figure~\ref{fig:4bus}). The line $l \in \mathcal{L}$
that connects bus $i$ and bus $j$ is denoted by $l = \{ i,j\}.$ The time of operation is denoted
by $t \in \RR.$

At bus $i,$ we denote the power generation and load at time $t$ by $G_{i,t}$ and $L_{i,t}$ respectively and  the reactance of link $l$ by $x_{l,t}.$ 
We adopt the dc power flow model \cite{wood1996power}, under which the power flow on line $l$ at time $t$ denoted by $F_{l,t},$ is given by $$F_{l,t} = \frac{1}{x_{l,t}}(\theta_{i,t} - \theta_{j,t}),$$ where $\theta_{i,t}$ and $\theta_{j,t}$ are the voltage phase angles at buses $i,j \in \mathcal{N}$ respectively at time $t$. For safe operation, the branch flows must be maintained within the power flow limits $F^{\max}_{k}$ at all time, i.e., $$-F^{\max}_{k} \leq F_{k,t} \leq F^{\max}_{k}, \ \forall t.$$
The relationship between branch power flows and the voltage phase angles can be compactly represented as $\fv_t = \Dm_t \Am^T \thetav_t,$ where the matrix $\Am \in \RR^{N \times L}$ is the branch-bus incidence matrix given by
\begin{align*}
\Am_{i,j} =
\begin{cases}
	1, & \text{if link $j$ starts at bus $i$},    \\
	-1, & \text{if link $j$ ends at bus $i$ }  , \\
	0 & \text{otherwise} ,
\end{cases} 
\end{align*}
and $\Dm_t \in \RR^{L \times L}$ is a diagonal matrix of the reciprocal of link reactances, i.e.,
\begin{align*}
\Dm_t = \text{diag} \LB \LSB \frac{1}{x_{1,t}},\frac{1}{x_{2,t}},\dots,\frac{1}{x_{L,t}} \RSB \RB,
\end{align*} 
and $\fv_t = [F_{1,t},\dots,F_{L,t}]^T$ (similarly $\gv_t, \lv_t, \thetav_t$ denote the vector forms of the corresponding quantities).

We assume that a subset of the links $\mathcal{L}_D \subseteq \mathcal{L}$ are equipped with D-FACTS devices, and the reactances of these links can be changed within the range $[\xv^{\min} , \xv^{\max}],$ where $\xv^{\min},\xv^{\max}$ are the reactance limits achievable by the D-FACTS devices. Naturally, $x^{\min}_l = x^{\max}_l = x_{l,t}$ if $l \notin \mathcal{L}_D.$ Denote the vector of branch reactances by $\xv_t.$

\subsection*{State Estimation \& Bad Data Detection Technique}
SE is a technique of estimating the system state from its noisy sensor measurements \cite{wood1996power}. Under the dc power flow model, the state at time $t$ corresponds to the nodal voltage phase angles $\thetav_t$, which are monitored by a set of $M$ measurements ${\zv}_t \in \RR^{M }.$ The measurements correspond to the nodal power injections, and the forward and reverse branch power flows, i.e. ${\zv}_t = [\tilde{\pv}_t,\tilde{\fv}_t,-\tilde{\fv}_t]^T.$ We note that the measurements may be different from the actual values of $\pv_t$ and $\fv_t$ due to sensor measurement noises or cyber-attacks. The measurement vector and the state are related as $${\zv}_t = \Hm_t \thetav_t + \nv_t,$$ where $\nv_t$ is the measurement noise, which is assumed to have Gaussian distribution. $\Hm_t \in \RR^{M \times N}$ is the measurement matrix given by
\begin{equation*}
\Hm = \LSB
\begin{array}{c}
\Dm_t \Am^T \\
-\Dm_t \Am^T \\
\Am \Dm_t \Am^T \\
\end{array} \RSB. 
\end{equation*}
The estimate of the system state, $\widehat{\thetav_t},$ is computed using a maximum likelihood (ML) estimation technique, given by \cite{wood1996power},
$$\widehat{\thetav_t} = ({\Hm^T_t} \Wm \Hm_t)^{-1} {\Hm^T_t} \Wm \zv_t,$$
where $\Wm$ is a diagonal weighting matrix whose elements are reciprocals
of the variances of the sensor measurement noise.

A BDD is used to detect faulty sensor measurements. It compares the residual defined by $r_t = ||\zv_t - \Hm_t \widehat{\thetav_t}||$ against a pre-defined threshold $\tau$
and raises an alarm if $r_t \geq \tau.$ 
The detection threshold $\tau$ is determined by the SO to ensure a certain false positive (FP) rate
$\alpha,$ where $\alpha > 0$ (usually a small value close to zero).

\subsection*{Undetectable FDI Attacks}
We consider FDI attacks against the SE, in which
the attacker injects an attack vector $\av_t \in \RR^{M}$ into the sensor measurements, 
i.e., $\zv^a_{t} = \zv_t+\av_t,$ where $\zv^a_{t}$ is the measurement vector under an attack. 
In general, the BDD can detect arbitrary FDI attack vectors. However, it is demonstrated \cite{Liu2009} that the BDD's detection probability for attacks of the form $\av_t = \Hm_t \cv,$ where $\cv \in \RR^N,$ is no greater than the FP rate $\alpha$. Such attacks are referred to as \emph{undetectable attacks}. 

\subsection*{Optimal Power Flow Problem}
OPF is an optimization framework to adjust the power flows in the network (by setting the generator dispatch and the branch reactances) with the objective of minimizing the generation cost for a given load vector $\lv_t \in \RR^{N},$ stated as follows\footnote{In the absence of D-FACTS devices installed within the grid, OPF optimizes over the generator dispatch values only (which is the version of OPF traditionally used \cite{wood1996power}).}:
\begin{subequations}
\label{eqn:OPF_normal}
\beqa
 C_{\text{OPF},t}  =  & \dsp \min_{ \gv_t,\xv_t} &  \sum_{i \in \mathcal{N}} C_i (G_{i,t}) \label{eqn:OPF_normala}
    \\ 
& s.t. &  \gv_t - \lv_t  = \Bm_t \thetav_t, \label{eqn:OPF_normalb}
 \\
& & -\fv^{\max} \leq \fv_t \leq \fv^{\max}, \label{eqn:OPF_normale}\\
& & \gv^{\min} \leq \gv_t \leq \gv^{\max}, \\
& & \xv^{\min} \leq \xv_t \leq  \xv^{\max}  \label{eqn:OPF_normald},
\eeqa 
\end{subequations}
where $C_i(G_{i,t})$ is the cost of generating $G_{i,t}$ units of power
at node $i \in \mathcal{N},$ the matrix $\Bm_t =  \Am \Dm_t \Am^T.$
In \eqref{eqn:OPF_normal}, the first constraint \eqref{eqn:OPF_normalb} represents the nodal power balance constraint, i.e., the power injected into a node must be equal to the power flowing out of the node. Constraints \eqref{eqn:OPF_normale}-\eqref{eqn:OPF_normald} correspond to the branch power flows, generator limits, and D-FACTS limits, respectively. We denote $\gv_t^*, \xv_t^* = \argmax_{\gv_t,\xv_t} \text{OPF}.$ We note that the OPF cost depends on the branch reactances through the matrix $\Bm_t$ (in addition to the loads).

\section{Moving-Target Defense in Power Grids}
\label{sec:MTD}
\begin{figure}[!t]
\centering
\includegraphics[width=0.45\textwidth]{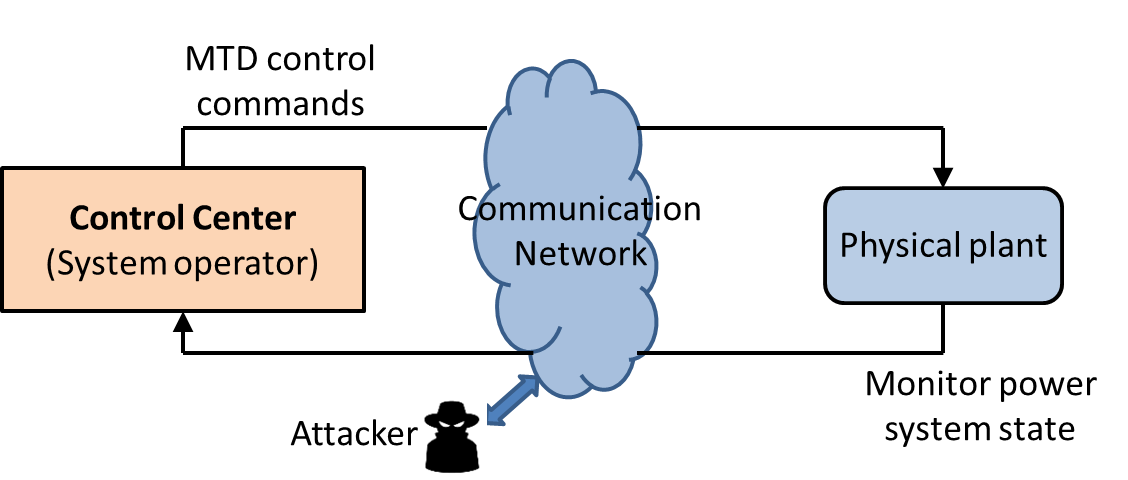}
\caption{System block diagram.}
\label{fig:MTD_System}
\end{figure}
\subsection{Attacker and the Defender Model}
\label{sec:AttackDefModel}
A block diagram of the system under study is shown in Fig.~\ref{fig:MTD_System}.
We consider a strong attacker who has access to the measurement data communicated 
between the field devices and the control center. Such access could be obtained by 
exploiting vulnerabilities in power grid communication systems. For example, in modern-day power grids, the field devices (such as remote terminal units) are often IP-accessible \cite{Shodan}. We also assume that the attacker can learn the system's measurement matrix (using the eavesdropped measurements) and craft undetectable FDI attacks accordingly (e.g., see \cite{KimTong2015,PoorBlind2013}).

Under MTD, the defender (e.g., the SO) tries to thwart the FDI attacks by actively perturbing the transmission line reactances to invalidate the attacker's prior knowledge. 
We assume that at the time of introducing MTD perturbations, there are no on-going FDI attacks. Note that the power system under consideration is naturally dynamic (even without MTD) since the branch reactances are optimized periodically to reflect temporal changes in the system load (refer to the OPF problem in \eqref{eqn:OPF_normal}). However, these natural changes are usually insufficient for effectively negating the attacker's knowledge. Thus, the defender deliberately introduces an additional reactance perturbation to ensure the MTD's detection capability.

The defender implements the MTD reactance perturbations by sending MTD control commands to the remote D-FACTS devices in the grid. Unlike the sensor measurements that support the grid's normal operation (e.g., extensive SCADA measurements collected every few seconds), these commands are much less frequent (e.g., hourly, see the discussion below), have much more restricted scope (i.e., between the control center and the set of D-FACTS devices only), and do not have stringent real-time constraints. Hence, we assume that it is feasible to encrypt the MTD commands to ensure their confidentiality. 

\begin{figure}
\begin{center}
     \begin{scriptsize}
       \def\svgwidth{ 0.95\columnwidth}
       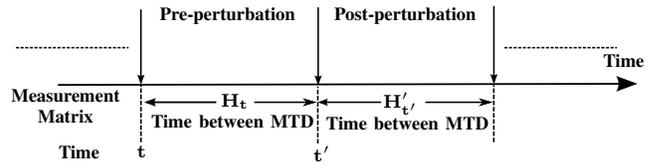
       \caption{MTD timeline. The vertical arrows indicate the times at which the system is perturbed.}
       \label{fig:Forecast}
       \end{scriptsize}
       \end{center}   
\end{figure}

We note that although the attacker cannot read the MTD commands directly due to their encryption, 
in principle he may still infer the MTD perturbations by monitoring their effects on the eavesdropped sensor measurements and estimating the new measurement matrix accordingly. Thus, the secrecy of the MTD generally decays over time. In practice, however, the learning will be time consuming since the attacker must collect an informative sequence of the measurements over a significant duration of time.  
In this paper, we assume that the time interval between the MTD perturbations is sufficiently small, so that during it the attacker's gain in knowledge (of the measurement matrix) is negligible. 

A guiding principle to estimate the perturbation time interval can be obtained from \cite{KimTong2015}, in which it is shown that FDI attacks against an IEEE 14-bus system require about $500-1000$ measurements of the system to successfully bypass the BDD, even if these measurements are assumed to have maximum information diversity in that they are i.i.d. 
Hence, if we assume optimistically for the attacker that SCADA measurements need to be only $5-10$ seconds apart to achieve the information diversity, their result suggests that the time required by the attacker to learn the system sufficiently well for stealthy attacks is on the order of a few hours. Accordingly, hourly MTD perturbations might be realistic for practical systems.
Further, we note that utilities typically solve the OPF more frequently, i.e., every $5-10$ minutes (whereas we only need to update the MTD every hour or so). Thus, between the MTD updates, the OPF will be solved as in \eqref{eqn:OPF_normal}.

The MTD timeline is illustrated in Fig.~\ref{fig:Forecast}. We consider two representative 
time instants $t$ and $t^\prime$ at which the reactances are perturbed for MTD. We denote the branch reactances and the measurement matrix after applying the MTD perturbations by $\xv^{\prime}_{t^\prime} = [x^{\prime}_{1,t^\prime},\dots,x^{\prime}_{L,t^\prime}]^T$ and ${\Hm}^{\prime}_{t^\prime}$ respectively, and the reactance perturbation vector by $\Delta \xv_{t,t^\prime} = \xv_t - \xv^{\prime}_{t^\prime}.$ We note that in the absence of MTD, the branch reactances and the measurement matrix would be set to $\xv_{t^\prime}$ and ${\Hm}_{t^\prime}$ by solving \eqref{eqn:OPF_normal} at time $t^\prime.$

In the rest of the paper, we address the question of how to select MTD perturbations that are effective in detecting FDI attacks crafted based on the outdated (i.e., pre-pertubation) knowledge, and examine
their cost-benefit tradeoff. 
We use $a^\prime_{t^\prime}$ to denote the value of a power system parameter $a_t$ after the MTD. E.g., $\theta^\prime_{t^\prime}$ denotes the nodal voltage phase angles after the MTD.  To motivate our inquisition, we now illustrate an example to show that certain randomly selected MTD perturbations will remain vulnerable to FDI attacks crafted with the attacker's pre-pertubation knowledge of the system. 

\begin{figure}[!t]
\centering
\includegraphics[width=0.47\textwidth]{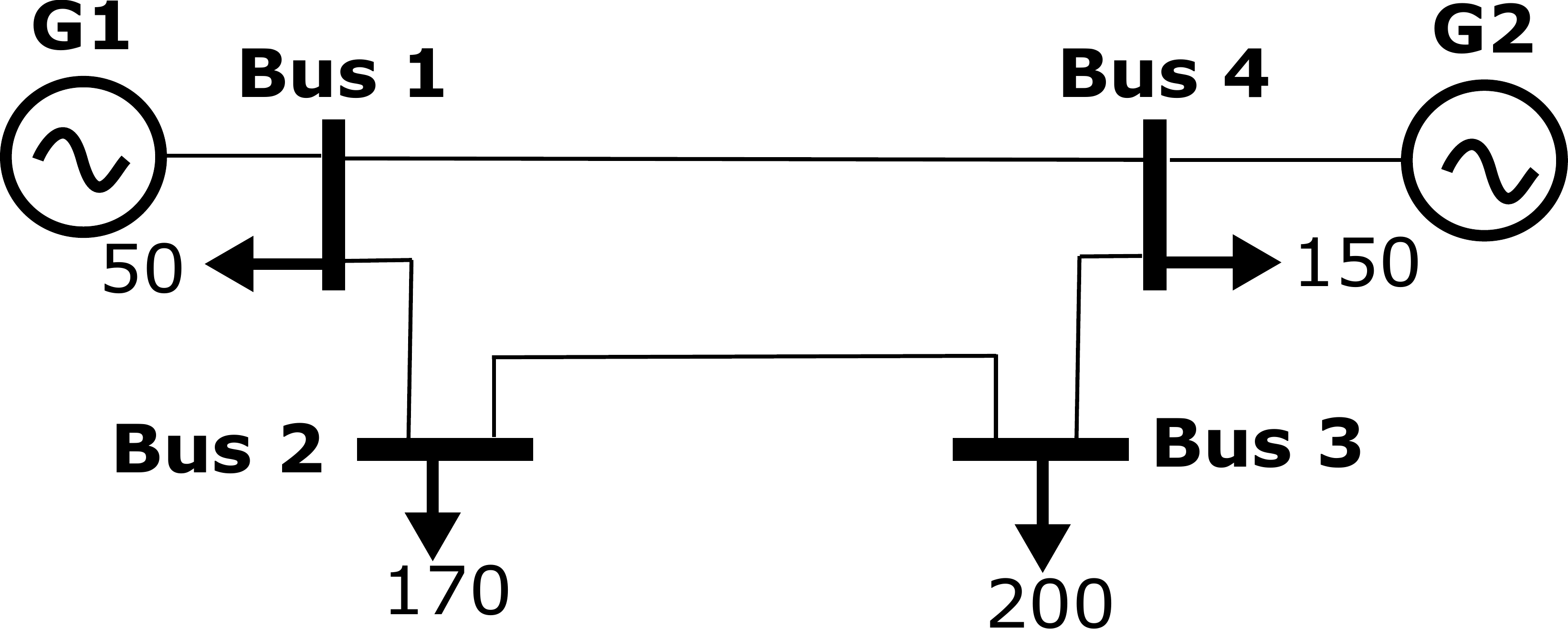}
\caption{$4$ bus system under consideration. The loads are indicated in MWs.}
\label{fig:4bus}
\end{figure}

\subsection{A Motivating Example} 
We consider the $4$-bus example shown in Fig.~\ref{fig:4bus} \cite{Matpower2011}. For simplicity, we assume that the system load is fixed (indicated in Fig.~\ref{fig:4bus}) and does not change with time. Furthermore, the pre-perturbation system state and the reactance settings $\xv_t$ (and $\Hm_t$) are adjusted by solving \eqref{eqn:OPF_normal}. The resulting branch flows, generation values and OPF cost are listed in Table~\ref{tbl:OPF_Values}. The attacker is assumed to have learned the pre-perturbation matrix $\Hm_t$.

To implement the MTD, we consider four reactance perturbation vectors respectively given by
$\Delta \xv^{(1)}_{t,t^\prime}  = \eta [x_1,0,0,0]^T, \  \Delta \xv^{(2)}_{t,t^\prime}  = \eta [0,x_2,0,0]^T, \Delta \xv^{(3)}_{t,t^\prime}  = \eta [0,0,x_3,0]^T, \ \Delta \xv^{(4)}_{t,t^\prime}  = \eta [0,0,0,x_4]^T,$ where $\eta$ is the percentage change in the reactance relative to its initial value. We assess each of the four MTD perturbations in terms of (i) attack detection and (ii) operational cost.

\begin{table}[!t]
\centering
\setlength\extrarowheight{4pt}
\begin{tabular}{|c|c|c|c|c|c|}
\hline
	 &  ${r^\prime}^{(1)}$ & ${r^\prime}^{(2)}$ & ${r^\prime}^{(3)}$ & ${r^\prime}^{(4)}$ \\ \hline 
	Attack $1$  & 2.82 & 2.87 & 0  & 0 \\ 
    \hline 
    Attack $2$ & 0 & 0 & 2.87 & 2.82 \\ 
    \hline
\end{tabular}
\caption{BDD residual values.}
\label{tbl:BDD_Res}
\end{table}

\begin{table}[!t]
\centering
\setlength\extrarowheight{4pt}
\begin{tabular}{|c|c|c|c|c|c|c|c|}
\hline
\multicolumn{4}{|c|}{Line Flow (MWs)} & \multicolumn{2}{|c|}{Gen.  (MWs)} & Cost(\$)  \\ \hline
    Line 1 & Line 2 & Line 3 & Line 4 & Gen 1 & Gen 2 & \multirow{2}{*}{\parbox{0.025\textwidth}{\center $1.15 \times 10^4$}}\\ \cline{1-6}
   126.56 & 173.44 & -43.44 &  -26.56 & 350 & 150  & \\
   \hline
\end{tabular}
\caption{Pre-perturbation power flows, generator dispatch and OPF cost for 4-bus system.}
\label{tbl:OPF_Values}
\end{table}

 \begin{table}[!t]
 \centering
\setlength\extrarowheight{4pt}
\begin{tabular}{|c|c|c|c|c|} \hline
   MTD &  \multicolumn{2}{|c|}{Gen. (MWs)} & OPF Cost (\$) \\ \hline
$\Delta \xv_{1}$  & $337.37$ & $162.62$ & $1.1626 \times 10^4$ \\ \hline
$\Delta \xv_{2}$  & $340.51$ & $159.48$ & $1.595 \times 10^4$  \\ \hline
$\Delta \xv_{3}$  & $348.62$ &  $151.37$ & $1.1514 \times 10^4$  \\ \hline
$\Delta \xv_{4}$   & $345.95$ &  $154.02$  & $1.154 \times 10^4$  \\
   \hline
\end{tabular}
\medskip
\caption{Post-perturbation generator dispatch and OPF cost.}
\label{tbl:Pflow_redispatch}
\end{table}

For attack detection, we inject an attack of the form $\av = \Hm_t \cv$ into the modified power network (after the MTD), and examine its BDD residual. For illustration, we consider two attacks -- attack~1
in which $\cv = [0,1,1,1]^T$ and attack~2 in which $\cv = [0,0,0,1]^T$ -- and set $\eta = 0.2.$ 
For simplicity, we ignore measurement noises. The BDD residuals under the four MTD perturbations are listed in Table~\ref{tbl:BDD_Res}. Note that in the absence of measurement noise, a non-zero value of the residual indicates the presence of attack. We observe that for each of the four perturbations, there exist attack vectors of the form $\av = \Hm_t \cv$,  which continue to bypass the BDD for the perturbed power network. 

We also enlist the post-pertubration OPF cost in Table~\ref{tbl:Pflow_redispatch}. We observe that the OPF cost increases in each of the four cases, compared to its pre-perturbation cost, and the perturbation $\Delta \xv_{3}$ incurs the least cost.

\subsection{MTD Perturbation Selection Challenges}
Based on the above illustrating example, we make the following conclusions. First, it is evident that a subset of the attacks of the form $\av = \Hm_t \cv$ continue to bypass the BDD after the MTD. Since the defender does not have prior knowledge of the actual attack vector (note that $\cv$ is chosen by the attacker), he cannot make an informed choice of which perturbation to adopt. Without such knowledge, the defender must select the MTD that is capable of detecting a largest subset of the possible attacks. The second design criterion is the MTD's operational cost, i.e., other things being equal, the defender prefers a least-cost MTD. In the following sections, we characterize formally the MTD's effectiveness and its operational cost, and present a framework for choosing appropriate MTD perturbations that balance between the two concerns.

\begin{figure*}[!t]
\centering
\includegraphics[width=0.85\textwidth]{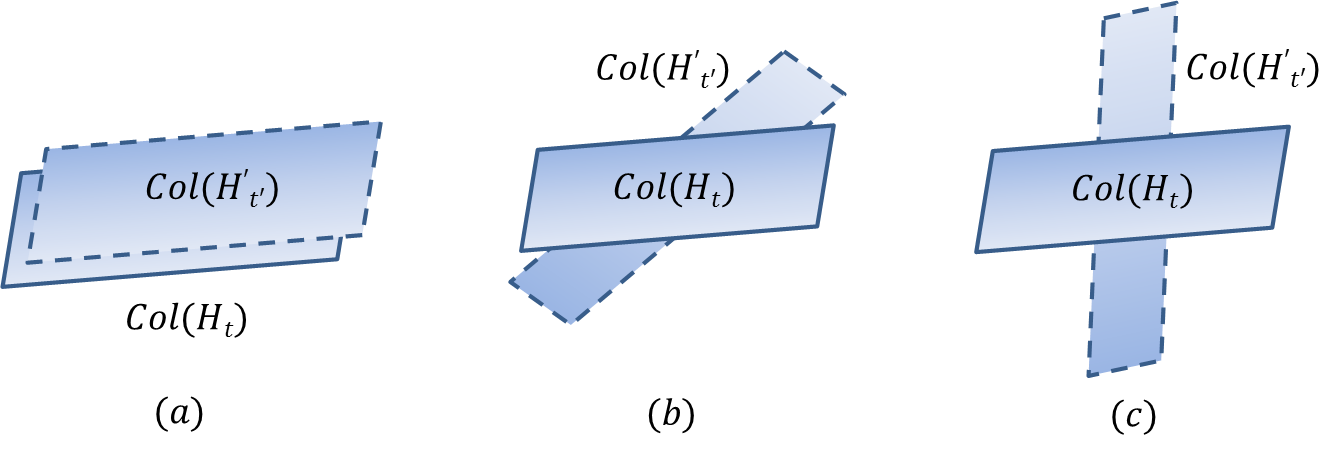}
\caption{Orientation of $Col(H^\prime_{t^\prime})$ with respect to $Col(H_t),$ (a) $\gamma(\Hm_t,\Hm^{\prime}_{t^\prime}) = 0$ (perfectly aligned column spaces), (b) $ 0 \leq \gamma(\Hm_t,\Hm^{\prime}_{t^\prime}) \leq \pi/2,$ and (c) $ \gamma(\Hm_t,\Hm^{\prime}_{t^\prime}) = \pi/2$ (orthogonal column spaces).}
\label{fig:Planes}
\end{figure*}

\section{MTD's Effectiveness of Attack Detection}
\label{sec:MTD_Att_Det}
In this section, we address the problem of selecting effective MTD reactance perturbations from an attack detection point of view. The goal is to select reactance perturbations within the physical constraints of the D-FACTS devices to effectively invalidate the attacker's knowledge for bypassing the BDD. 
The section is divided into two parts. In the first part, we devise a metric to quantify the effectiveness of the MTD. In the second part, we derive the conditions and propose design criteria for MTD perturbations to preclude stealthy FDI attacks in practice.

Henceforth, we use the notation ``MTD $\Hm^\prime_{t^\prime}$" to refer to a reactance perturbation that changes the measurement matrix from $\Hm_t$ to $\Hm^\prime_{t^\prime}.$ We let $\mathcal{A}$ denote the set of all attack vectors of the form $\av = \Hm_t \cv,$ i.e., 
$$\mathcal{A}  = \{ \av : \av = \Hm_t \cv, ||\av|| \leq a_{\max}, \cv \in \RR^{N} \}. $$
For an attack vector $\av,$ we let $P^{\prime}_D (\av)$ denote its detection probability under MTD $\Hm^\prime_{t^\prime},$ where $P^{\prime}_D (\av) = \mathbb{P} ( r^{\prime} \geq \tau).$ We denote by $\mathcal{A}^{\prime} (\delta) \subseteq \mathcal{A}$ 
the subset of attacks in $\mathcal{A}$ whose detection probability under MTD $\Hm^{\prime}_{t^\prime}$ is greater than a given $\delta \in [0,1],$ i.e.,
$$\mathcal{A}^{\prime} (\delta)  = \{ \av : \av = \Hm_t \cv,  ||\av|| \leq a_{\max}, P^{\prime}_D (\av) > \delta, \cv \in \RR^{N} \}.$$

\subsection{Metric to Quantify MTD's Effectiveness}
First, we devise a metric to quantify the MTD's effectiveness. Intuitively, an MTD perturbation ``A" is more effective than a perturbation ``B" if it can detect more FDI attacks in the set $\mathcal{A}$ with high probability. However, $\mathcal{A}$, a subset in the $n$-dimensional space ($\RR^n$), has infinitely many attack vectors. For these sets, the \emph{Lebesgue measure} generalizes the notion of length (one-dimensional), area (two-dimensional), or volume (three-dimensional) to $n$-dimensions \cite{tao2011mestheory}. The effectiveness of an MTD $\Hm^{\prime}_{t^\prime}$ for a given $\delta \in [0,1],$ which we denote by $\eta^\prime (\delta),$ can be quantified as
\begin{align}
\eta^\prime (\delta) = \frac{\lambda ( \mathcal{A}^\prime (\delta))} {\lambda (\mathcal{A})} , \label{eqn:eta}
\end{align}
where $\lambda ( \mathcal{A}^\prime (\delta))$ and $\lambda(\mathcal{A})$ denote the Lebesgue measures of the respective sets. Intuitively, $\eta^\prime (\delta)$ represents the ratio of the number of attack vectors of the form $\av = \Hm_t \cv$ whose detection probability under MTD $\Hm^{\prime}_{t^\prime}$ is greater than $\delta$ to the total number of attacks in the set $\mathcal{A}$. 
Since $\mathcal{A}^\prime (\delta) \subseteq \mathcal{A}$, $0 \leq \eta^\prime (\delta) \leq 1.$

Of particular interest are the sets $\mathcal{A}^\prime(\alpha)$ and $\mathcal{A} \setminus \mathcal{A}^\prime(\alpha),$ and the latter is the set of undetectable attacks under MTD $\Hm^\prime_{t^\prime}$ (refer to Section~\ref{sec:Prelim} for the definition of undetectable attacks). 
An ideal MTD is one that admits no undetectable attacks of the form $\av = \Hm_t \cv$, i.e., $\mathcal{A}^\prime(\alpha) = \mathcal{A}$ and $\eta^{\prime} (\alpha) = 1.$ In the following subsection, we derive  conditions on the MTD $\Hm^\prime_{t^\prime}$ that can ensure the property.

\subsection{MTD Admitting No Undetectable Attacks}
We start by characterizing the condition for an attack $\av = \Hm_t \cv$ to remain
undetectable under MTD $\Hm^{\prime}_{t^\prime}.$ 
\begin{proposition}
\label{lem:undetect_attacks}
An attack of the form $\av = \Hm_t \cv$ is \emph{undetectable} under MTD perturbation $\Hm^{\prime}_{t^\prime}$ if it satisfies the condition $\rank (\Hm^{\prime}_{t^\prime} ) = \rank ([\Hm^{\prime}_{t^\prime} \ \Hm_t \cv] ),$ where $[\Hm^{\prime}_{t^\prime} \ \Hm_t \cv] $ 
is the augmented matrix. 
\end{proposition}
The proof of this proposition is presented in Appendix~A. Intuitively, the proposition implies that an attack vector of the form $\av = \Hm_t \cv$ is undetectable under MTD $\Hm^{\prime}_{t^\prime}$ if it lies in the column spaces of both $\Hm_t$ and $\Hm^{\prime}_{t^\prime},$ 
since $\rank (\Hm^{\prime}_{t^\prime} ) = \rank ([\Hm^{\prime}_{t^\prime} \ \Hm_t \cv] )$ for the attack vector 
$\av = \Hm_t \cv \in Col(\Hm^{\prime}_{t^\prime}).$

The result allows us to give conditions for the MTD $\Hm^{\prime}_{t^\prime}$ to ensure no undetectable attacks of the form $\av = \Hm_t \cv.$ In particular, to achieve the aforementioned property, MTD $\Hm^{\prime}_{t^\prime}$ must be selected such that no attack vector $\av$ in the column space of $\Hm_t$ lies in the column space of $\Hm^{\prime}_{t^\prime}.$ The following theorem states the condition.

\begin{theorem}
\label{lem:orth_MTD}
An MTD $\Hm^{\prime}_{t^\prime}$ has no undetectable attacks of the form $\av = \Hm_t \cv$ if $Col(\Hm^{\prime}_{t^\prime})$ is the orthogonal complement of $Col(\Hm_t).$ Furthermore, for a given attack vector $\av$, such an MTD achieves the maximum value of $P^\prime_D (\av)$ among all the possible MTD perturbations. 
\end{theorem}
The proof is presented in Appendix~B. The first statement of this theorem implies that 
for the MTD $\Hm^\prime_{t^\prime}$ satisfying the orthogonality condition, there are no attacks 
of the form $\av = \Hm_t \cv$ for which $P^\prime_D (\av)$ is as low as the FP rate $\alpha$ (in general, $\alpha$ is chosen by the SO to be a small value). 
However, this result does not automatically imply that the attacks will also be detected with high probability, which is the desired outcome. But
the second statement of Theorem~\ref{lem:orth_MTD} shows that this is indeed the case, since such an MTD also maximizes $P^\prime_D (\av)$ among all possible MTD perturbations.

From Theorem~\ref{lem:orth_MTD}, we conclude that purely from an attack detection point of view, an MTD perturbation should be selected to achieve the stated orthogonality condition. However, this may not always be feasible due to practical limitations, e.g., the D-FACTS devices may only allow the reactances to be perturbed within a certain range. In these cases, we require an additional design criterion to select the MTD perturbations, which is the subject of the following subsection.

\subsection{Heuristic Design Criteria for Selecting MTD Perturbation}
\label{sec:Heuristic}
Intuitively, if the reactance adjustment capability of D-FACTS is insufficient to meet the orthogonality condition of Theorem~\ref{lem:orth_MTD}, the MTD perturbation should be selected to make $Col(\Hm^{\prime}_{t^\prime})$ as orthogonal to $Col(\Hm_t)$ as possible within the constraints of the D-FACTS device. To formalize this notion, we introduce the concept of \emph{principal angle} 
between subspaces, defined as follows:
\begin{defn}[\cite{AngSubspace1971}]
The smallest principal angle (SPA) $0 \leq \theta \leq \pi/2$  between the subspaces $\mathcal{F}, \mathcal{G}  \subseteq \CC^N$ is defined as $$
\cos (\theta)  = \dsp  \max_{ \substack {\uv \in \mathcal{F},\uv \in \mathcal{G} \\ ||\uv|| = 1, ||\vv|| = 1}} |\uv^H \vv|. $$
\end{defn}
The SPA generalizes the concept of angle between a pair of vectors to a pair of $n$-dimensional subspaces. 
Let $\gamma(\Hm_t,\Hm^{\prime}_{t^\prime})$ denote the SPA between $Col(\Hm_t)$ and $Col(\Hm^{\prime}_{t^\prime}).$ We conjecture that MTD perturbations with a higher value of $\gamma(\Hm_t,\Hm^{\prime}_{t^\prime})$ are more effective in terms of attack detection. Thus, SPA can be utilized as a design criterion for selecting good MTD perturbations.

The conjecture is based upon the following observations. (i) In Appendix~C, we present arguments which suggest that the attack detection probability $P^\prime_D (\av)$ increases as we select MTD perturbations with higher $\gamma(\Hm_t,\Hm^{\prime}_{t^\prime}).$ (ii) In the following, we give some observations to suggest that the measure of the set of undetectable attacks decreases by selecting MTD perturbations with higher $\gamma(\Hm_t,\Hm^{\prime}_{t^\prime}).$ 

We examine MTD perturbations in two extreme cases as illustrated in Fig.~\ref{fig:Planes}. First, consider MTD $\Hm^{\prime}_{t^\prime} = (1+ \eta) \Hm_t,$ for which it can be verified that $\gamma(\Hm_t,\Hm^{\prime}_{t^\prime}) = 0.$ For such an MTD, the column spaces of the matrices $\Hm_t$ and $\Hm^{\prime}_{t^\prime}$ are perfectly aligned. Hence all attacks of the form $\av = \Hm_t \cv$ remain undetectable after the MTD (i.e., $\mathcal{A}^\prime (\alpha) = \emptyset $ and $\lambda ( \mathcal{A}^\prime (\alpha)) = 0$). Thus, an MTD perturbation with $\gamma(\Hm_t,\Hm^{\prime}_{t^\prime}) = 0$ is the least effective in detecting FDI attacks. 
Second, for MTD $\Hm^{\prime}_{t^\prime}$ satisfying 
the orthogonality condition of Theorem~\ref{lem:orth_MTD}, it can be verified that $\gamma(\Hm_t,\Hm^{\prime}_{t^\prime}) = \pi/2.$ As shown in the previous subsection, in this case, $\mathcal{A}^\prime (\alpha) = \mathcal{A}$ and there are no undetectable attacks of the form $\av = \Hm \cv.$

These arguments suggest that MTD perturbations for which $\gamma(\Hm_t,\Hm^{\prime}_{t^\prime})$ is closer to $\pi/2$ are more effective in detecting FDI attacks, a trend that is also confirmed by our simulation results using the IEEE 14-bus system (see Section~\ref{sec:Sim_Res}). A natural follow up question is how to select the reactance perturbation vector $\Delta \xv_{t,t^\prime}$ to achieve the aforementioned design criteria. In the next section, we present an optimization framework to numerically compute $\Delta \xv_{t,t^\prime}$ while also considering the MTD's operational cost.

\section{MTD's Cost-Benefit Tradeoff} 
\label{sec:MTD_Cost}
Thus far, we have investigated the MTD from an 
attack detection point of view only. In this section, we formally define the operational
cost of MTD in an optimization framework.

\subsection*{MTD Operational Cost} 
We quantify MTD's cost in terms of the 
increase in OPF cost due to the MTD relative to its value without MTD, i.e., 
\begin{align}
C_{\text{MTD},t^\prime} = \frac{ {C^\prime_{\text{OPF},t^\prime}} -   C_{\text{OPF},t^\prime}}{C_{\text{OPF},t^\prime}}, \label{eqn:C_MTD}
\end{align}
where $C_{\text{OPF},t^\prime}$ is the OPF cost of the system corresponding to 
the measurement matrix $\Hm_{t^\prime}$ computed using \eqref{eqn:OPF_normal} (at time $t^\prime$), and $C^\prime_{\text{OPF},t^\prime}$ is the OPF cost of the system with MTD (corresponding to 
the measurement matrix $\Hm^\prime_{t^\prime}$). Note that $C_{\text{MTD},t^\prime}$ is always
non-negative since the additional perturbation due to MTD will increase the OPF cost.

From \eqref{eqn:C_MTD}, we note that $C_{\text{MTD},t^\prime}$ depends on the separation between the 
column spaces of $\Hm_{t^\prime}$ and $\Hm^\prime_{t^\prime}$. In particular, if the two matrices are identical, then $C_{\text{MTD},t^\prime}$ is zero. As the separation between the column spaces of the two matrices $\gamma(\Hm_{t^\prime},\Hm^\prime_{t^\prime})$ is increased, the power flows within the two systems and the corresponding generation dispatch will be different (due to the reactance perturbation). Consequently, the OPF cost in the system with MTD perturbation will increase. 

Our observation is that $\gamma(\Hm_t,\Hm^\prime_{t^\prime})$ closely approximates 
$\gamma(\Hm_{t^\prime},\Hm^\prime_{t^\prime}).$ Hence, MTD's operational cost increases as we choose perturbations with higher $\gamma(\Hm_t,\Hm^\prime_{t^\prime}).$ The approximation can be explained as follows. Recall that $\Hm_{t}$ and $\Hm_{t^\prime}$ differ only due to temporal variations in the system load. Since the power system load is temporally correlated, the matrices $\Hm_t$ and $\Hm_{t^\prime}$ will not differ significantly and their column spaces are nearly aligned. Thus, $\gamma(\Hm_t,\Hm^\prime_{t^\prime})$ can be used as an approximate measure of the SPA between the column spaces of $\Hm_{t^\prime}$ and $\Hm^\prime_{t^\prime}$. Extensive simulation results driven by real-world data load traces presented in Section~\ref{sec:Sim_Res} confirm the validity of this approximation.

\subsection*{MTD Tradeoff} 
Following the above arguments, we note that the defender faces conflicting objectives. On the one hand, for the MTD to be effective from an attack detection point of view, the column spaces of the matrices $\Hm_t$ and $\Hm^{\prime}_{t^\prime}$ should be 
as orthogonal as possible. On the other hand, the MTD's operational cost increases with $\gamma(\Hm_t,\Hm^{\prime}_{t^\prime})$. Thus, there exists a trade-off between the MTD's effectiveness and its operational cost. To balance the two aspects, we formulate the MTD reactance selection problem as a constrained optimization problem with the objective of minimizing the operational cost subject to a constraint on the MTD's effectiveness. The problem is stated as: 
\begin{subequations}
\label{eqn:OPF_perturb}
\beqa
C^\prime_{\text{OPF},t^\prime} = &\dsp \min_{ \gv^\prime_{t^\prime}, \xv^\prime_{t^\prime}} &  \sum_{i \in \mathcal{N}} C_i (G^\prime_{i,t^\prime}) \label{eqn:OPF_perturb-a}
    \\ 
& s.t. &  \gamma(\Hm_t,\Hm^{\prime}_{t^\prime}) \geq \gamma_{\text{th}}, \label{eqn:OPF_perturb-b}
\\
& & \gv^\prime_{t^\prime} - \lv_{t^\prime}  = \Bm^{\prime}_{t^\prime} \thetav^{\prime}_{t^\prime}, \label{eqn:OPF_perturb-c}  \\ 
& & -\fv^{\max} \leq \fv^\prime_{t^\prime} \leq \fv^{\max}, \\
& & \gv^{\min} \leq \gv^\prime_{t^\prime} \leq \gv^{\max}, \\
& & \xv^{\min} \leq \xv^\prime_{t^\prime} \leq  \xv^{\max} .
\eeqa 
\end{subequations}
In \eqref{eqn:OPF_perturb}, the SPA between the column spaces of $\Hm_t$ and $\Hm^\prime_{t^\prime}$ is used as a heuristic metric to approximate the effectiveness of the attack detection $\eta^\prime(\delta)$  (based on the conjecture stated in Section~\ref{sec:MTD_Att_Det}-C). 
In \eqref{eqn:OPF_perturb-b}, we impose a constraint on the SPA, where 
$\gamma_{\text{th}} \in [0,\pi/2]$ is a threshold that must be tuned numerically (see Section~\ref{sec:Sim_Res} for more details). Simulation results show that different values of the threshold $\gamma_{\text{th}}$ provide a spectrum of trade-offs between the MTD's effectiveness and its operational cost. We propose to solve \eqref{eqn:OPF_perturb} numerically using existing constrained non-linear optimization solvers (e.g., the \emph{fmincon} function of MATLAB).

Note that the attacker does not have sufficient information to solve \eqref{eqn:OPF_perturb} and thus cannot anticipate the MTD perturbations. In particular, at time $t^\prime$, the attacker does not know $\Hm_t$, since there is not sufficient time to learn it given the frequency of perturbations (see the discussion in Sec.~\ref{sec:AttackDefModel}). Hence, the secrecy of the MTD is satisfied.

\section{Simulation Results}
\label{sec:Sim_Res}
In this section, we present simulation results to evaluate the MTD's 
effectiveness and its operational cost.

\subsection{Simulation Settings \& Methodology}
The simulations are carried out in MATLAB. All the constrained optimization problems involved in the simulations are solved using the \emph{fmincon} function of MATLAB with the \emph{MultiStart} algorithm.

\begin{figure}[!t]
\centering
\includegraphics[width=0.5\textwidth]{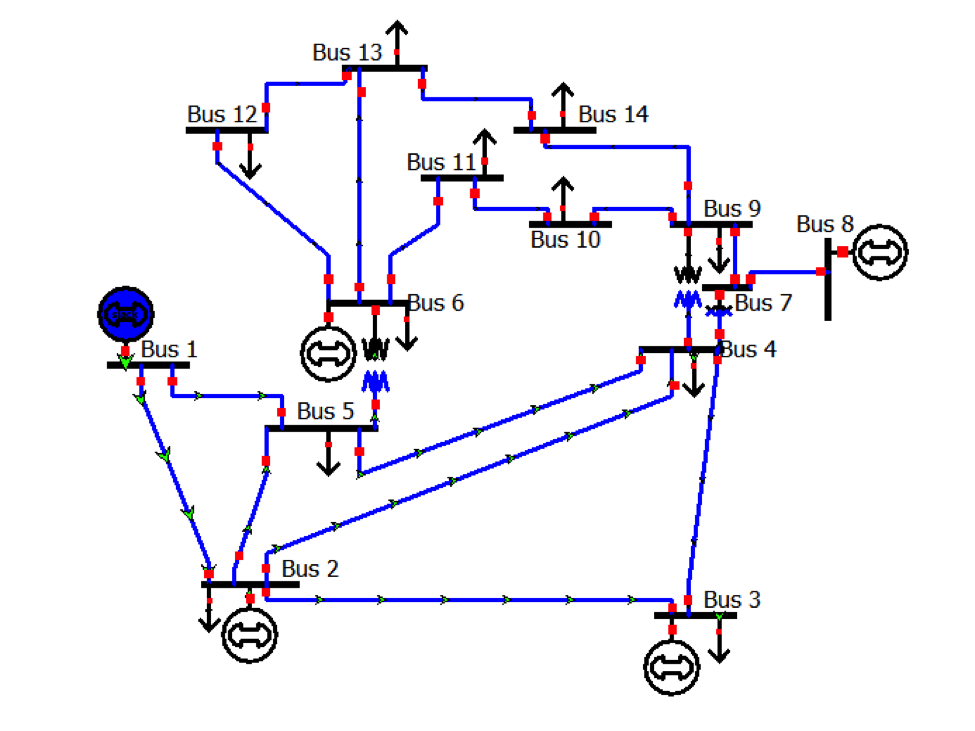}
\caption{IEEE 14-bus system. (Figure source: \cite{IEEE14})}
\label{fig:14bus}
\end{figure}

We perform simulations using the IEEE 14-bus system. The bus topology is shown in Fig.~\ref{fig:14bus}. We obtain its configuration data from the MATPOWER package \cite{Matpower2011}. 
As shown in Fig.~\ref{fig:14bus}, the generators are installed at buses $1,2,3,6,8$ and
their parameters are listed in Table~\ref{tbl:Gen_Limits}.
We use the linear generation cost model given by $C_i (G_{i,t}) = c_i G_{i,t}.$ 
We assume that D-FACTS devices are installed on 
$6$ branches indexed by $\mathcal{L}_D = \{ 1,5,9,11,17,19\}.$ 
The D-FACTS limits are set to $\xv_{\min} = (1-\eta_{\max}) \xv$ and $\xv_{\min} = (1+\eta_{\max}) \xv,$ where $\xv$ is the default values (obtained from the IEEE~14-bus case file) and $\eta_{\max}$ is set to $0.5$.
Further, the branch flow limits are chosen to be $160$ MWs for link $1,$ and $60$ MWs for all other links of the power system. The rest of the settings are obtained from the MATPOWER configuration case file.

\begin{table}
\centering
\setlength\extrarowheight{4pt}
\caption{Generator parameters.}
\begin{tabular}{llllll}
\hline
 Gen. bus & 1 &  2 & 3 & 6 & 8  \\ \hline
$P_{\max}$ (MWs) &   300 & 50 & 30 &  50 &  20 \\ \hline
$c_i$ (\$/MWh) &   20 & 30 & 40 &  50 & 35 \\
   \hline
\end{tabular}
\label{tbl:Gen_Limits}
\end{table}

\subsection{Simulation Results with Static Load}
In the first set of simulations, we assume that the system load is static (we use default values from the IEEE 14-bus MATPOWER case file). The pre-perturbation reactances $\xv_t$ (and $\Hm_t$) are adjusted by solving \eqref{eqn:OPF_normal}. The defender designs MTD $\Hm^\prime_{t^\prime}$ assuming that the attacker has acquired the knowledge of $\Hm_t,$ and that he injects attacks of the form $\av = \Hm_t \cv.$

\begin{figure*}[!t]
\centering
\begin{subfigure}{0.5\textwidth}
\includegraphics[width=1\textwidth]{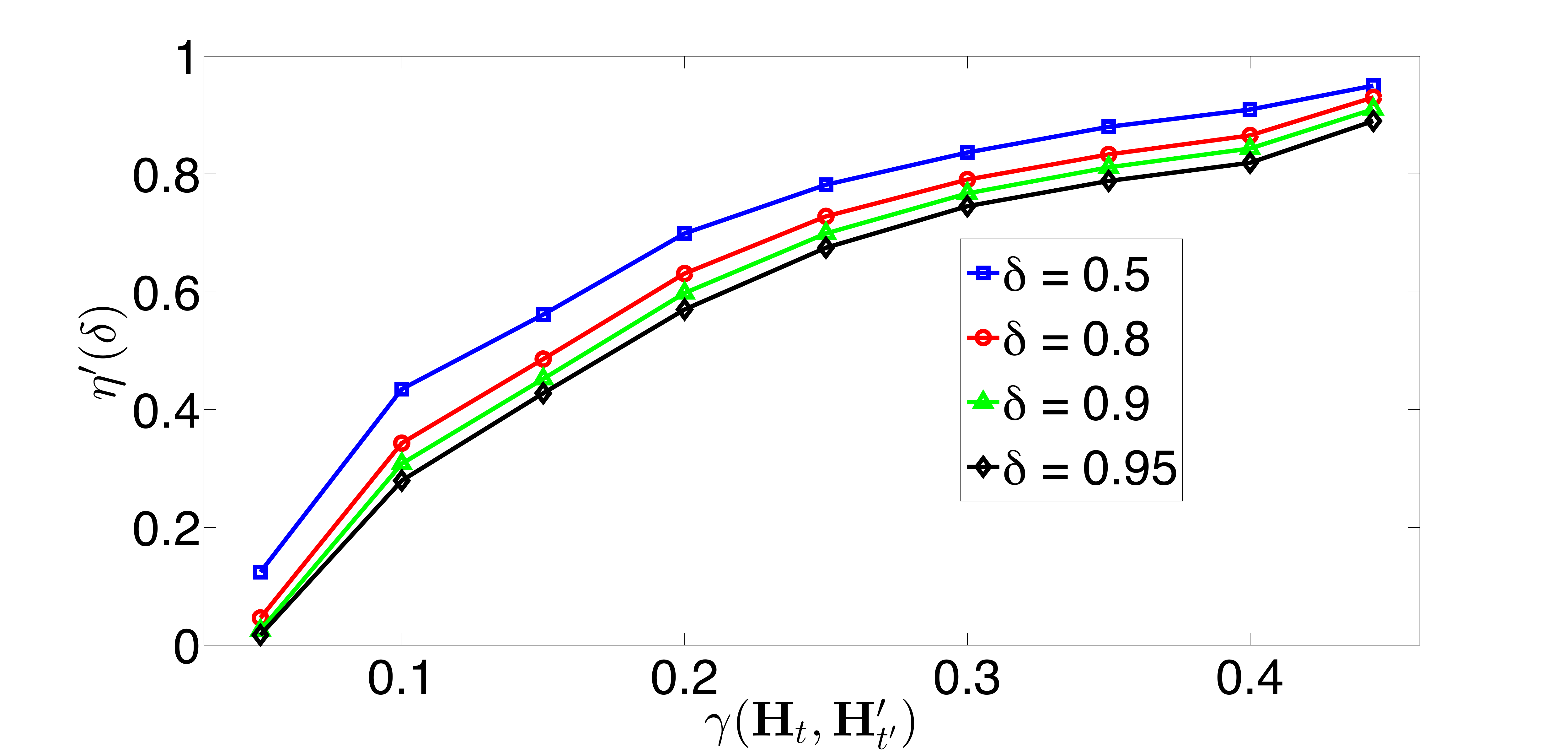}
\caption{IEEE 14-Bus System}
\end{subfigure}
~
\begin{subfigure}{0.48\textwidth}
\includegraphics[width=0.96\textwidth]{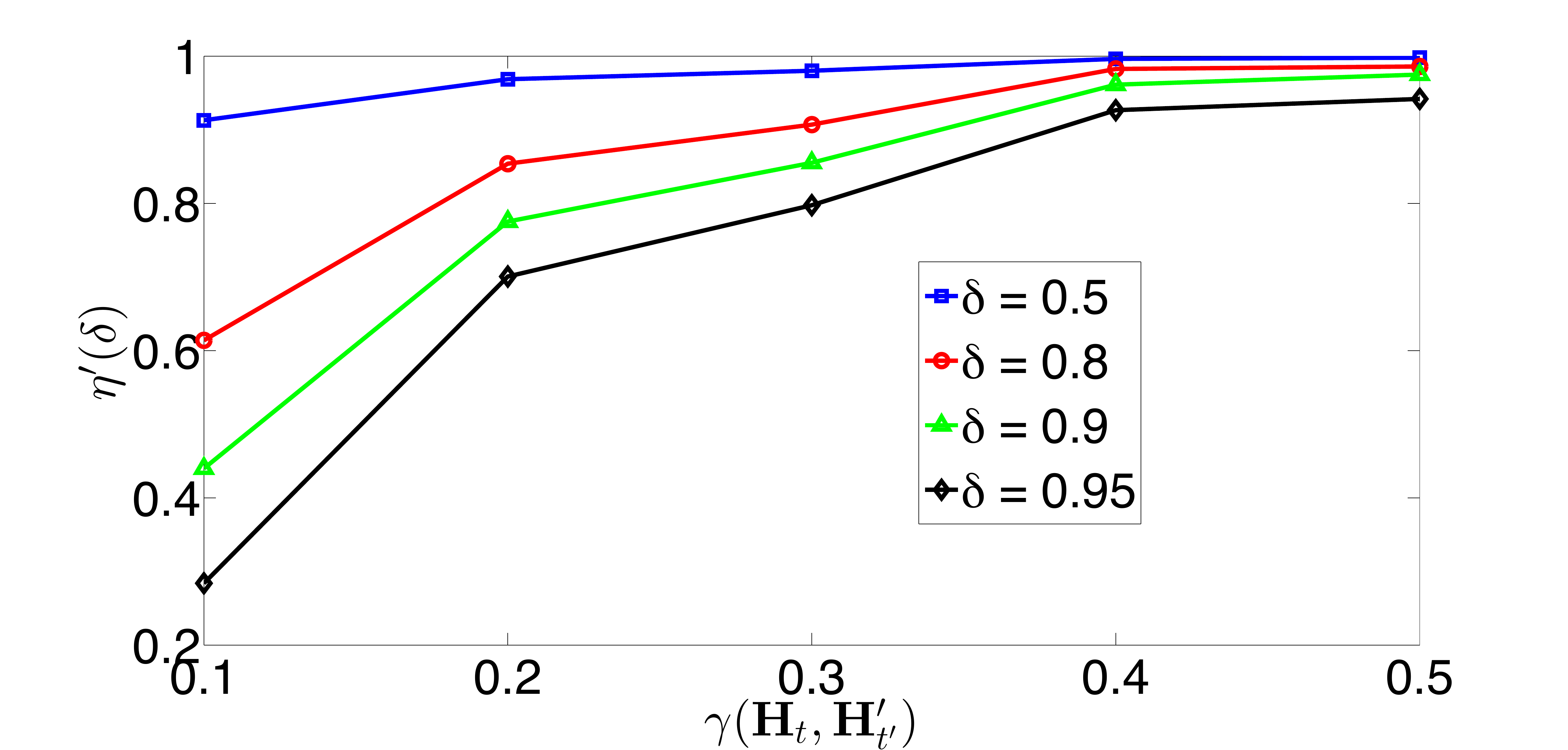}
\caption{IEEE 30-Bus System}
\end{subfigure} 
\caption{MTD effectiveness for different values of $\gamma (\Hm_t,\Hm^{\prime}_{t^\prime})$ (radians). FP rate is set to $5 \times 10^{-4}.$}
\label{fig:MTD_Det}
\end{figure*}

\subsubsection*{Effectiveness of Attack Detection} 
First, we examine the MTD's effectiveness ($\eta^\prime(\delta)$) for different values of $\gamma(\Hm,\Hm^{\prime}).$ We choose $\gamma(\Hm_t,\Hm^{\prime}_{t^\prime}) \in [0,0.45]$ radians in steps of $0.05$ radians. For each value of $\gamma(\Hm_t,\Hm^{\prime}_{t^\prime})$, we solve the optimization problem \eqref{eqn:OPF_perturb} by setting $\gamma_{\text{th}}$ to the corresponding value, and evaluate $\eta^\prime(\delta)$ using Monte Carlo simulations as follows.  We consider 
$1000$ attack vectors of the form $\av = \Hm_t \cv,$ where the vector $\cv$ is chosen as a random vector drawn from the Gaussian distribution, and scale its magnitude such that $||\av||_1/||\zv||_1 \approx 0.08$ (the scaling adjusts the magnitude of attack injections to be relatively small in comparison to the actual measurements). 
We then evaluate $P^{\prime}_D (\av)$ for each of the attack vectors (the details will be presented shortly), and count the fraction of attack vectors for which $P^{\prime}_D (\av) \geq \delta,$ for a given value of $\delta \in [0,1].$ 
For each attack vector, the detection probability $P^{\prime}_D (\av)$ is computed by 
generating $1000$ instantiations of measurement noise (according to the Gaussian distribution), and counting the number of times the BDD alarm is triggered. The BDD threshold is adjusted such that the FP rate is set to $5 \times 10^{-4}.$ We note that MTD does not alter the FP rate of the BDD.

In Fig.~\ref{fig:MTD_Det} (a), we plot the variation of $\eta^\prime(\delta)$ as a function 
of $\gamma(\Hm_t,\Hm^{\prime}_{t^\prime})$ for different values of $\delta$. 
In this figure, the y-axis represents the fraction of attacks for which $P^{\prime}_D (\av) \geq \delta,$ for a given  $\gamma(\Hm_t,\Hm^{\prime}_{t^\prime})$. We observe that $\eta^\prime(\delta)$ monotonically increases with 
$\gamma(\Hm_t,\Hm^{\prime}_{t^\prime}),$ thus confirming our intuition that MTD perturbations with higher values of $\gamma(\Hm_t,\Hm^{\prime}_{t^\prime})$ are more effective in attack detection. E.g., for $\gamma = 0.44,$ $97 \%$ of the attacks have a detection probability greater than $0.95.$ In practice, the defender can run these simulations to determine an appropriate $\gamma_{\text{th}}$ for meeting a desired level of attack detection.

\subsubsection*{Comparison With Existing Work}
\begin{figure}[!t]
\centering
\includegraphics[width=0.5\textwidth]{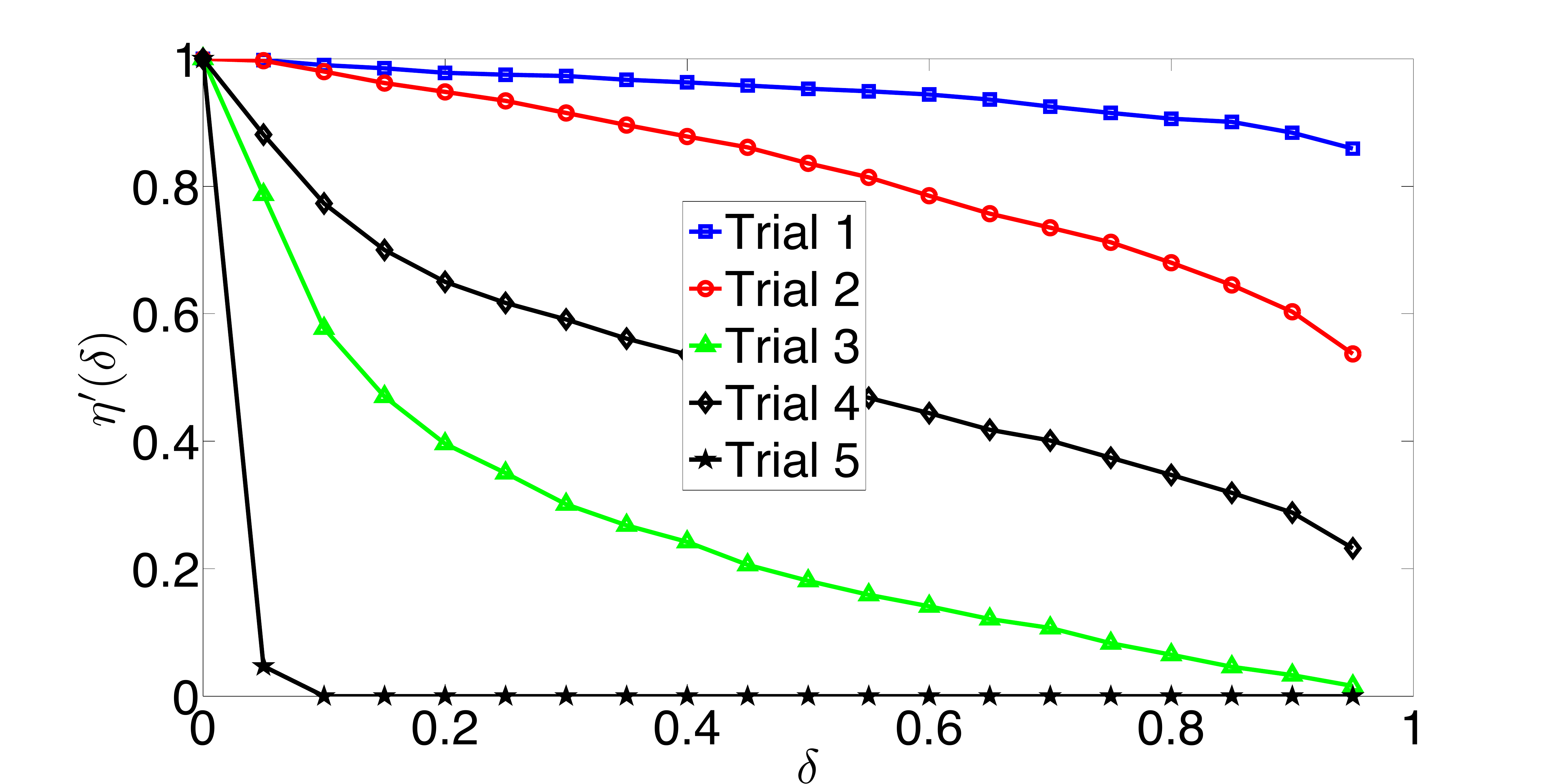}
\caption{MTD effectiveness under five randomly chosen MTD perturbations in IEEE 14-bus system. FP rate is set to $5 \times 10^{-4}.$}
\label{fig:MTD_Random_Det}
\end{figure}

We also perform simulations to compare our MTD selection approach with
state of the art \cite{Morrow2012,Davis2012,RahmanMTD2014}. 
Similar to the related work, we implement MTD by selecting random MTD perturbations that are 
constrained to be within $2\%$ of the optimal value.  
We plot $\eta^{\prime} (\delta)$ as a function of $\delta$ for five such randomly-chosen perturbations in Fig.~\ref{fig:MTD_Random_Det}. It can be seen that $\eta^{\prime} (\delta)$ exhibits high variability across the trials, implying that the randomly chosen MTD perturbations
cannot always guarantee effective attack detection. 

\begin{figure}[!t]
\centering
\includegraphics[width=0.5\textwidth]{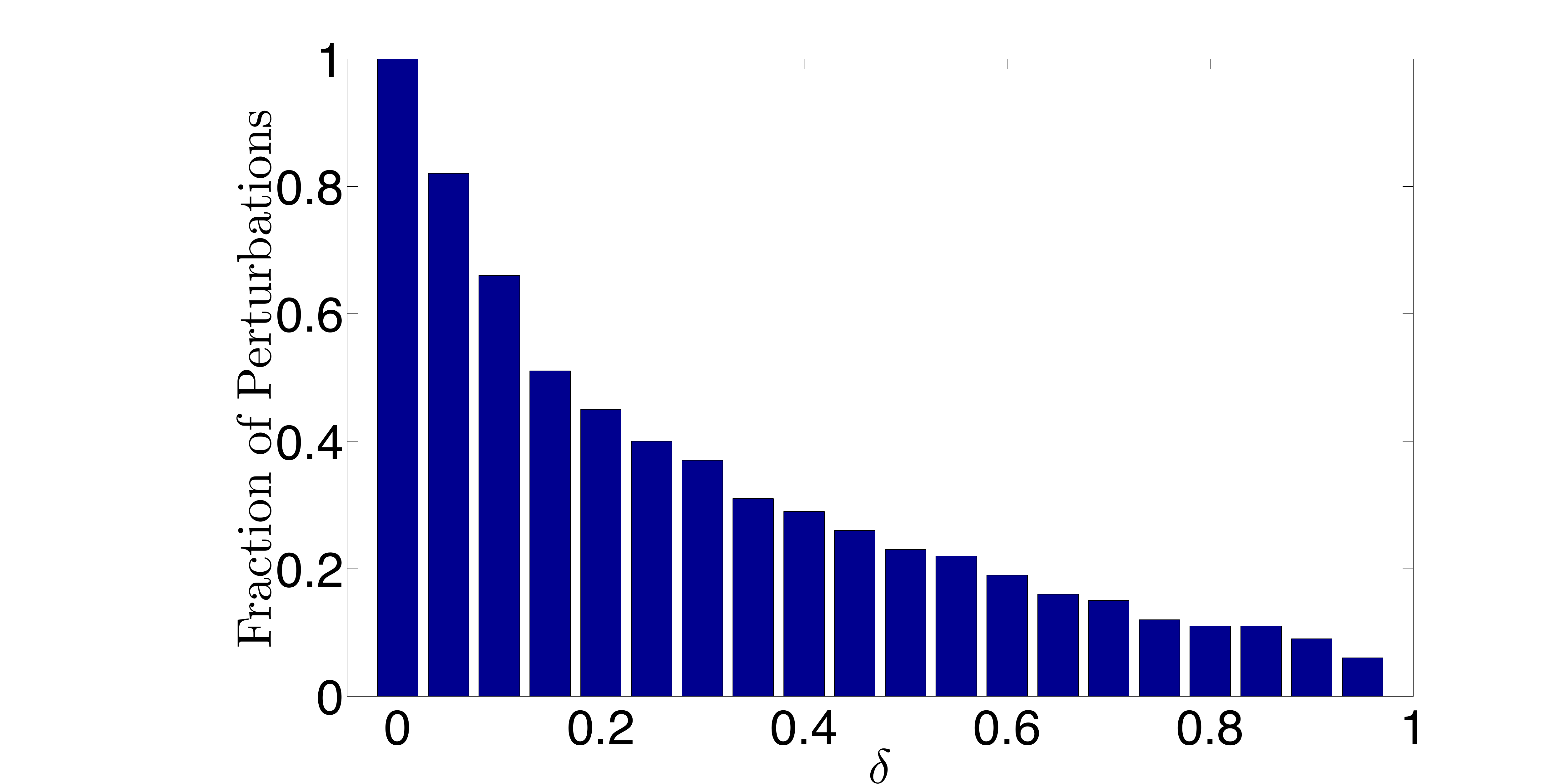}
\caption{Fraction of randomly-chosen MTD perturbations that satisfy $\eta^\prime (\delta) \geq 0.9.$}
\label{fig:Random_MTD_Fraction}
\end{figure}

Further, out of $500$ such randomly chosen perturbations (known also as the \emph{keyspace} \cite{Morrow2012,Davis2012}), we count the fraction of perturbations which satisfy $\eta^\prime (\delta) \geq 0.9$ for different values of $\delta,$ and plot the results in Fig.~\ref{fig:Random_MTD_Fraction}. We observe that less $10 \%$ of the randomly-selected MTD perturbations satisfy $\eta^\prime (0.9) \geq 0.9.$ 
In contrast, the MTD perturbations chosen according to our approach can always guarantee a certain effectiveness, once the subspace angle threshold $\gamma_{\text{th}}$ is adjusted to an appropriate value. This highlights the importance of designing the MTD according to the formal design criterion advanced in this work.

To show the scalability of the proposed approach to larger bus systems, we plot the $\eta^\prime(\delta)$ as a function of $\gamma (\Hm_t,\Hm^{\prime}_{t^\prime})$ for 
the IEEE 30-bus system in Fig.~\ref{fig:MTD_Det} (b). We use default settings provided in the MATPOWER case file. We observe results similar to those for the IEEE 14-bus system, i.e., perturbations which have a higher value of $\gamma (\Hm_t,\Hm^{\prime}_{t^\prime})$ are more effective in terms of attack detection.

\subsection{Simulation Results With Dynamic Load} 
In the next set of simulations, we consider dynamic load. We use a load data trace from New York state for one day (25-JAN-2016) \cite{NYISO} sampled hourly, and feed it to the IEEE 14-bus system. The simulations are performed every hour. 
At each hour, $C_{\text{OPF},t}$ is computed by solving \eqref{eqn:OPF_normal} with the load input of the corresponding hour. On the other hand, $C^\prime_{\text{OPF},t^\prime}$ is computed by solving \eqref{eqn:OPF_perturb} assuming that the attacker's knowledge is outdated by $1$~hour. For example, while computing the MTD $\Hm^\prime_{t^\prime}$ at $9$~AM, we assume that the attacker has acquired the knowledge of the measurement matrix $\Hm_t$ at $8$~AM. (Recall from our previous discussion in Sec.~IV-A that hourly MTD perturbations are realistic for practical systems.)

\begin{figure}[!t]
\centering
\includegraphics[width=0.5\textwidth]{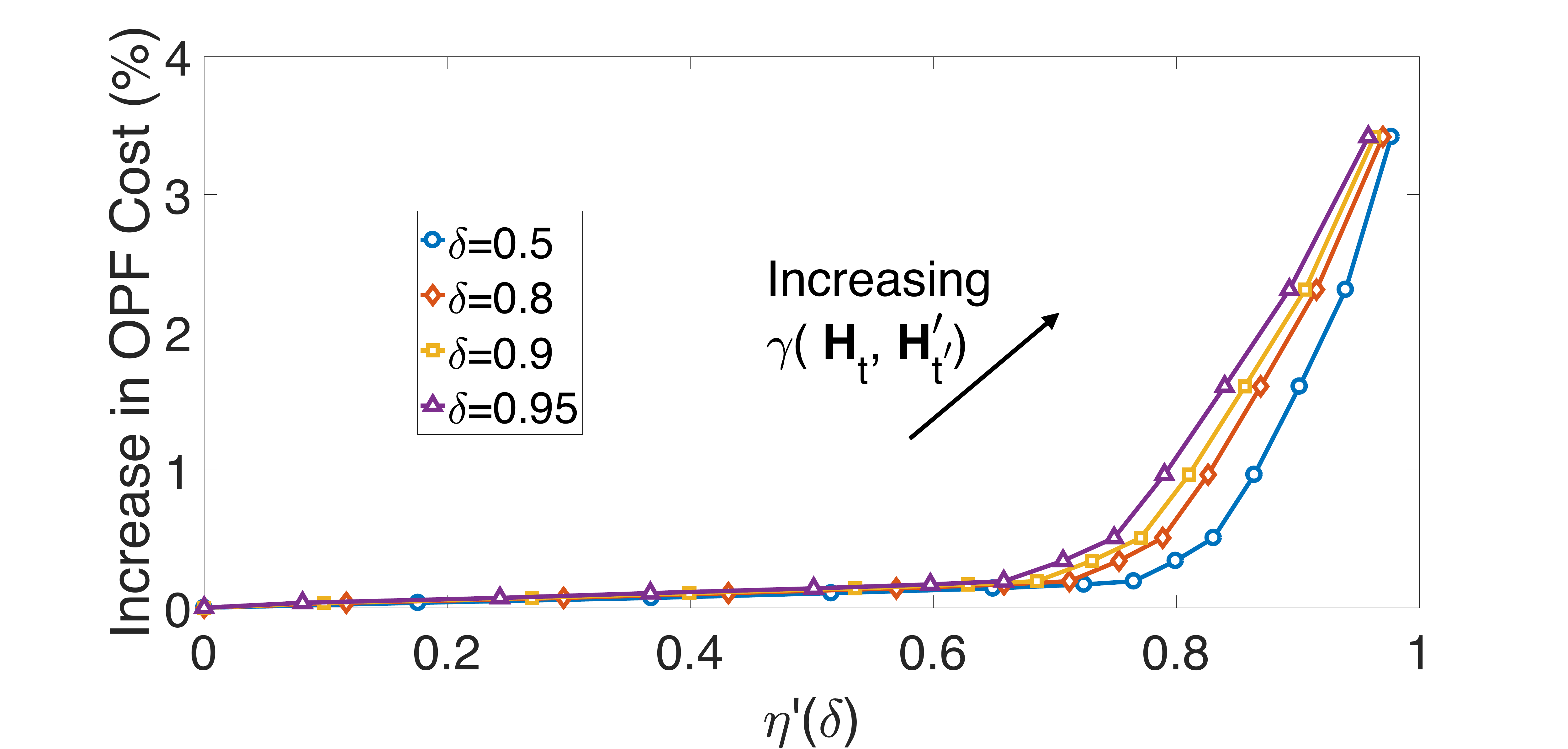}
\caption{Tradeoff between MTD's effectiveness and operational cost in IEEE 14-bus system. The data corresponds to $6$~PM.}
\label{fig:Trade_Off_Curve}
\end{figure}

\subsubsection*{MTD Tradeoff}
In Fig.~\ref{fig:Trade_Off_Curve}, we plot of the tradeoff between $\eta^{\prime} (\delta)$
and the operational cost for data corresponding to $6$~PM.
We make the following observations. For low values of $\eta^{\prime} (\delta),$ the operational cost is nearly zero. However, as $\gamma(\Hm_t,\Hm^{\prime}_{t^\prime})$ and consequently $\eta^{\prime} (\delta)$ is increased, the MTD incurs a non-trivial operational cost. 
In particular, the cost increases steeply for values of $\eta^{\prime} (\delta)$ very close to $1.$ 
E.g., for $\delta = 0.9,$ an increase in the value of $\eta^{\prime}(\delta)$ from $0.8$ to 
$0.9$ changes the MTD operational cost from $0.96\%$ to $2.31\%.$ 
These results suggest that the defender must carefully choose an appropriate level of attack detection while taking into account the increase in operational cost.

\subsubsection*{MTD Operational Cost Over a Day} 
\begin{figure}[!t]
\centering
\includegraphics[width=0.5\textwidth]{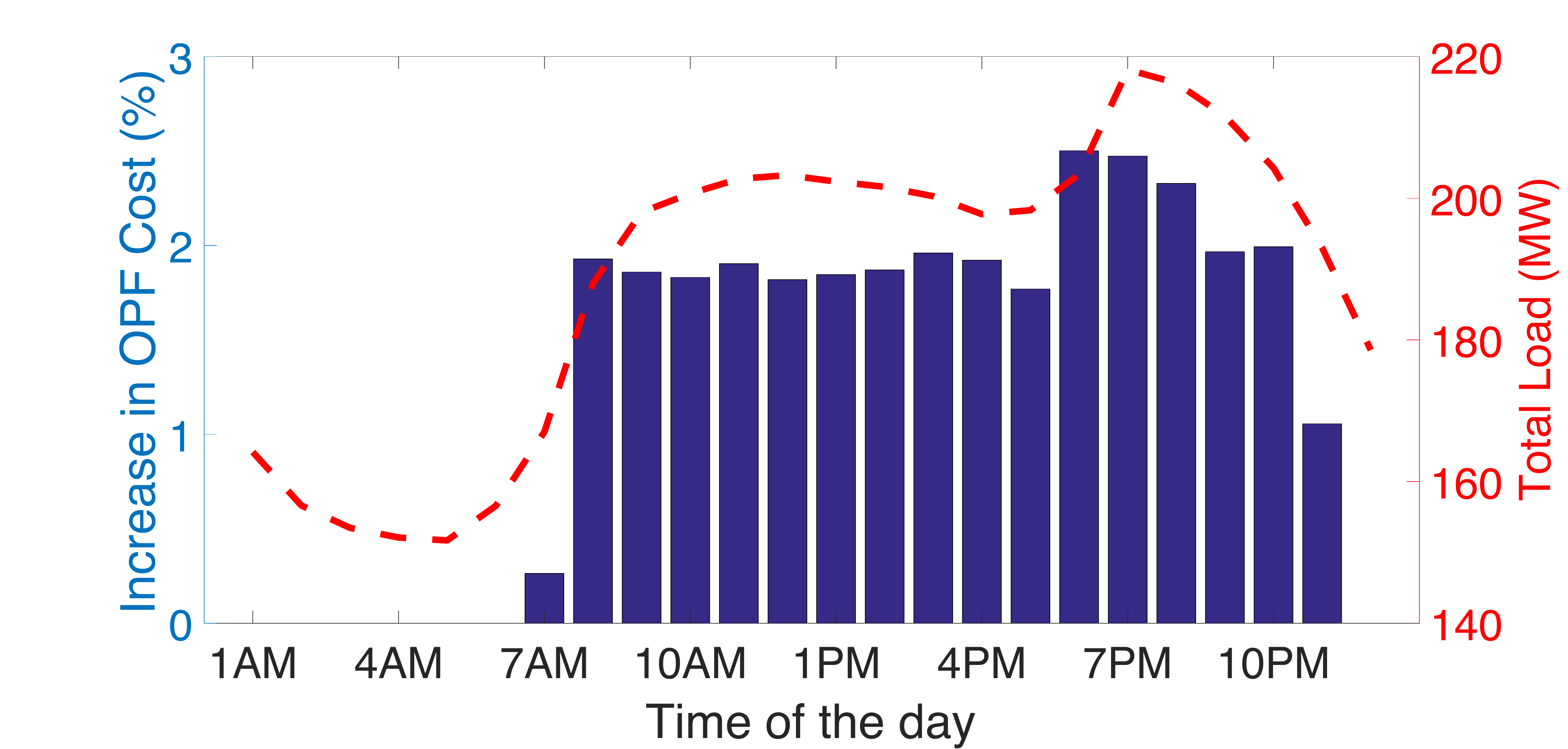}
\caption{MTD operational cost over a day computed using New York state hourly load data trace (25-JAN-2016).}
\label{fig:NY_Load}
\end{figure}

\begin{figure}[!t]
\centering
\includegraphics[width=0.5\textwidth]{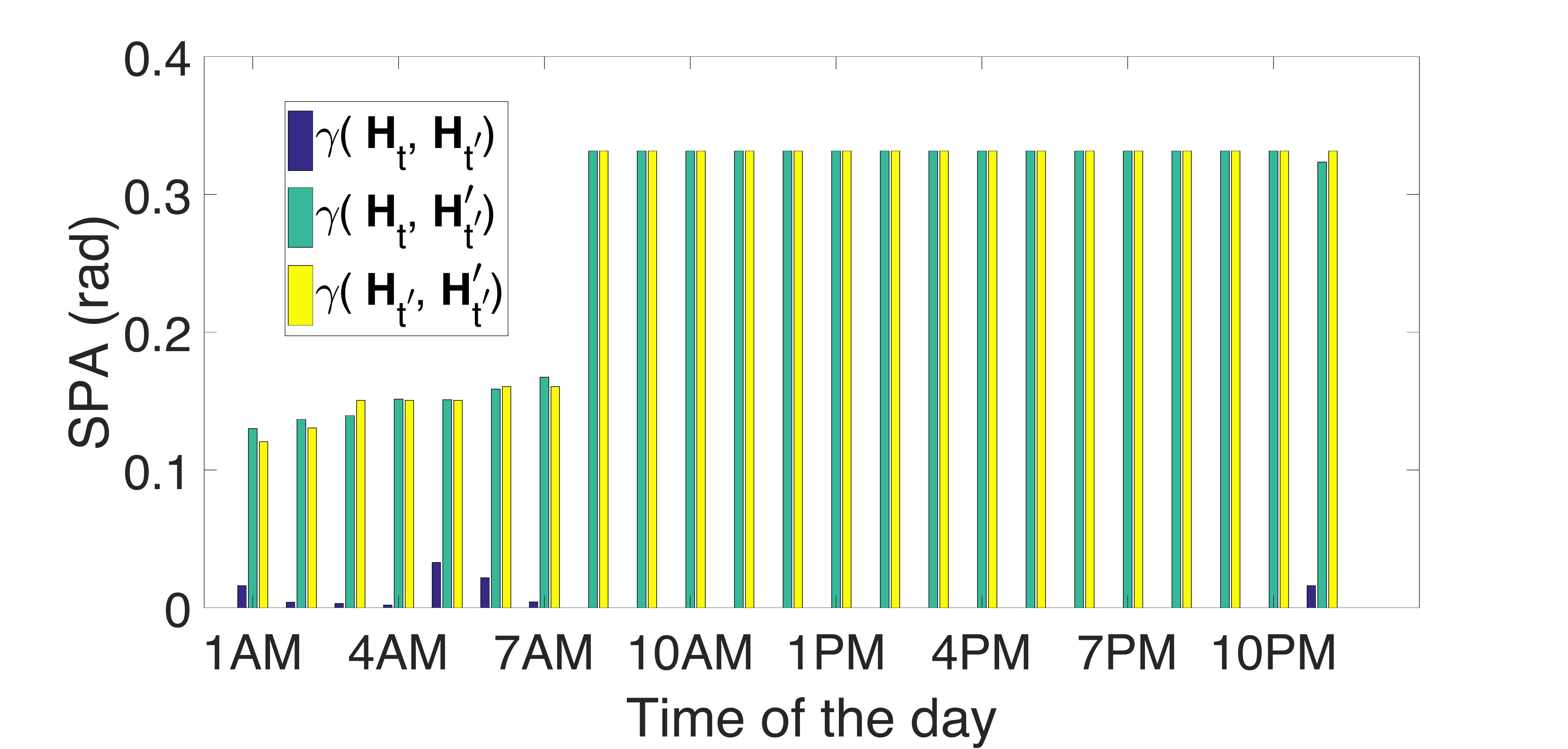}
\caption{Smallest principal angle (in radians) between pre-perturbation and post-perturbation measurement matrices.}
\label{fig:Angles}
\end{figure}

We also perform simulations to show how the cost varies over the day. At each hour, we adjust the subspace angle threshold $\gamma_{\text{th}}$ numerically such that the MTD perturbation achieves effectiveness of $\eta^{\prime} (0.9) \geq 0.9.$ The corresponding value of $\gamma(\Hm_{t^\prime},\Hm^\prime_{t^\prime})$ is shown in Fig.~\ref{fig:Angles}. The rest of the bus settings is identical to the previous simulation. The variation of MTD operational cost and the aggregate load are shown in Fig.~\ref{fig:NY_Load}. It can be observed that the MTD operational cost increases at higher load. This can be explained as follows. 
When the system load is low, there will be a significant buffer capacity between the branch power flows and the corresponding flow limits. If the difference in power flows between the two systems (with and without MTD) is within the buffer capacity, then the generator dispatch in the two systems will be identical (or close to each other). Thus, the corresponding MTD cost is low. At higher loads, the power system is significantly congested, and the branch power flows of the two systems (with and without MTD) will differ significantly. Consequently the generator dispatch in the two systems will be different  leading to an increase in the OPF cost.

We also plot the quantities $\gamma(\Hm_t,\Hm_{t^\prime})$ and $\gamma(\Hm_{t^\prime},\Hm^\prime_{t^\prime})$ for every hour in Fig.~\ref{fig:Angles}. We observe that $\gamma(\Hm_t,\Hm_{t^\prime})$ is nearly zero for all the simulation instants. This is because the matrices $\Hm_t$ and $\Hm_{t^\prime}$ do not differ significantly due to the temporal correlation of the system load between different simulation instants and their column spaces are nearly aligned.  These results also validate the approximation $\gamma(\Hm_t,\Hm^\prime_{t^\prime}) \approx \gamma(\Hm_{t^\prime},\Hm^\prime_{t^\prime}).$

\subsection{Discussion}
\label{sec:Prac_Cons}
To put the MTD operational cost in perspective, we can compare it against the potential cost of damage due to a BDD-bypassing attack. For example, prior work \cite{RenLRAttacksTPDS2012,RenLoadRedis2011} suggests that such an attack can increase the OPF cost by up to $28 \%,$ and additionally cause transmission line trips (considering IEEE 14-bus system with similar simulation settings). Our numbers suggest that the MTD's operational cost is comparatively significantly smaller. In practice, based on its own deployment scenario and other factors like estimated likelihood of attacks, the SO can make similar comparisons to assess the merits of adopting the MTD defense.

\section{Conclusions}
\label{sec:Conc}
We addressed the problem of selecting MTD reactance perturbations that
are truly effective in thwarting stealthy FDI attacks against SE in power grids. We devised a novel metric to quantify the MTD's effectiveness, and identified key design criteria to compute effective MTD perturbations in practice. We also showed that the effective MTD may incur a non-trivial operational cost, and provided analysis to expose the cost-benefit tradeoff of the MTD in an OPF framework. Our result offers MTD to system operators as an insurance against possible FDI attacks, and minimizes the cost of such insurance subject to an effectiveness constraint.

\balance
\bibliographystyle{IEEEtran}
\bibliography{IEEEabrv,bibliography}

\section*{Appendix A: Proof of Proposition~\ref{lem:undetect_attacks}}
To simplify notation, in this appendix, we drop the time subscripts $t$ and $t^\prime$
from the relevant quantities. 

A sketch of the proof is as follows. First, we express the residual 
$r^\prime $ as the sum of two components, a noise component $\rv^\prime_n$ and an 
attack component  $\rv^\prime_a,$ given by $ r^\prime = ||\rv^\prime_n + \rv^\prime_a ||.$ We then show that for attacks that satisfy the condition of Proposition~\ref{lem:undetect_attacks},  $\rv^\prime_a = \bf{0},$ and hence their detection probability is no greater than the FP rate.

We proceed with the first step of the proof. Recall the expression of $r^\prime =
||{\zv}^{\prime} - {\Hm}^{\prime} \widehat{{\thetav}^{\prime}}||,$ where ${\zv}^{\prime} = {\Hm}^{\prime} \thetav^{\prime} + \nv + \Hm \cv,$ $\widehat{{\thetav}^{\prime}} = ({\Hm^{\prime}}^T \Wm \Hm^{\prime})^{-1} {\Hm^{\prime}}^T \Wm {\zv}^{\prime}.$ It can be simplified as
\begin{align}
 r^\prime & = ||\zv^{\prime}  - \Hm^{\prime} ({\Hm^{\prime}}^T \Wm \Hm^{\prime})^{-1} {\Hm^{\prime}}^T \Wm \zv^{\prime}|| \nonumber \\ 
& = ||\Hm^{\prime} {{\thetav}^{\prime}} + \nv + \Hm \cv \nonumber \\ & \qquad \qquad - \Hm^{\prime} ({\Hm^{\prime}}^T \Wm \Hm^{\prime})^{-1} {\Hm^{\prime}}^T \Wm (\Hm^{\prime} {{\thetav}^{\prime}}  + \nv + \Hm \cv)||  \nonumber \\
& = || (\Id - \Gammam^{\prime}) \nv + (\Id - \Gammam^{\prime}) \Hm \cv||, 
\end{align}
where $\Gammam^{\prime} = \Hm^{\prime} ({\Hm^{\prime}}^T \Wm \Hm^{\prime})^{-1} {\Hm^{\prime}}^T \Wm.$ 
We note that $r^\prime$ consists of two components, a noise component $\rv^\prime_n \defines (\Id - \Gammam^{\prime}) \nv,$ and an attack component $\rv^\prime_a \defines (\Id - \Gammam^{\prime}) \Hm \cv.$ If $\rv^\prime_a = {\bf 0},$ then the detection probability of $\av$ is no greater than the FP rate $\alpha$, and hence, the attack is undetectable under the MTD perturbation $\Hm^{\prime}$. 
Note that for all the attacks $\av = \Hm \cv \in Col(\Hm^{\prime}),$ $\rv^\prime_a = {\bf 0}.$ In other words, 
the system of equations $\Hm \cv = \Hm^{\prime} \cv^{\prime}$ must be consistent, for some
$\cv^{\prime } \in \RR^N.$ This condition holds true if and only if $\rank (\Hm^{\prime} ) = \rank ([\Hm^{\prime} \ \Hm \cv] )$ \cite{Meyer:2000}.

\section*{Appendix~B: Proof of Theorem~\ref{lem:orth_MTD}}

A sketch of the proof is as follows. We prove the first statement by showing that for an MTD $\Hm^{\prime}$ satisfying the orthogonality condition, $\rv^\prime_a = \bf{0}$ if an only if $\cv = \bf{0}.$ Thus it follows that there are no non-zero attacks that are undetectable under such an MTD. 
To prove the second statement, we show that $P^\prime_D (\av)$ increases as we increase $|| \rv^\prime_a ||.$ Furthermore, we show that $|| \rv^\prime_a ||$ achieves its maximum value under the MTD perturbation that satisfies the conditions of Theorem~\ref{lem:orth_MTD}.

We begin with the proof of the first statement of Theorem~\ref{lem:orth_MTD}. 
If $Col(\Hm^{\prime})$ is the orthogonal complement of $Col(\Hm),$ then ${\Hm^{\prime}}^T  \Wm \Hm \cv = {\bf 0}, \ \forall \cv \in \RR^{N},$ since $\Hm \cv \in Col(\Hm).$ In this case, $\rv^\prime_a$ becomes
\begin{align}
\rv^\prime_a  = \Hm \cv - \Hm^{\prime} ({\Hm^{\prime}}^T \Wm \Hm^{\prime})^{-1} {\Hm^{\prime}}^T \Wm  \Hm \cv \nonumber = \Hm \cv. \label{eqn:residue_orth}
\end{align}
Recall that an attack is undetectable if $\rv^\prime_a = {\bf 0}.$ For MTD $\Hm^{\prime}$ that satisfies the orthogonality condition, 
substituting for $\rv^\prime_a$ from \eqref{eqn:residue_orth}, we have that $\Hm \cv = {\bf 0}.$
Since $\Hm$ is a full rank matrix, the set of equations $\Hm \cv = {\bf 0}$ has a unique solution $\cv = {\bf 0}$ \cite{Meyer:2000}. 
Hence, there are no non-zero undetectable attacks of the form $\av = \Hm \cv.$

Next, we prove the second statement of Theorem~\ref{lem:orth_MTD}. First, note that under any MTD $\Hm^{\prime},$ $||\rv^\prime_a||$ can be bounded as $0 \leq ||\rv^{\prime}_a|| \leq ||\av||.$ The lower bound is true in a straightforward manner. The upper bound follows from 
\begin{align}
||\rv^{\prime}_a|| = ||(\Id - \Gammam^{\prime}) \av|| \leq ||(\Id - \Gammam^{\prime})|| \ || \av|| = ||\av||, 
\end{align}
where the last equality is due to the fact that $\Id - \Gammam^{\prime}$ is a projection matrix and hence has unit norm.
Furthermore, under any MTD $\Hm^{\prime},$ $r^{\prime} = ||\rv^\prime_n + \rv^\prime_a||$ follows a \emph{noncentral chi-square} distribution \cite{Muirhead1982} with its noncentrality parameter equal to $||\rv^\prime_a||$ (since  $\rv^\prime_n + \rv^\prime_a$ 
is a Gaussian random variable with $\rv^\prime_a$ as its mean). 

For a  non-central chi-square distributed random variable $X$, $\mathbb{P} ( X \geq \tau )$ increases by increasing the noncentrality parameter. Hence, we can conclude that the quantity $P^{\prime}_D (\av) = \mathbb{P} ( r^\prime \geq \tau )$ increases
by increasing $||\rv^\prime_a||$. For an attack vector $\av,$ the quantity $||\rv^\prime_a||$ depends on the choice of MTD $\Hm^\prime.$
Thus, we can conclude that MTD perturbations that yield a greater value of $||\rv^\prime_a||$ can detect the attack vector $\av$ with 
higher probability (i.e., $P^{\prime}_D (\av)$ is higher).

In particular, for MTD $\Hm^{\prime}$ that satisfies the conditions of Theorem~\ref{lem:orth_MTD}, from \eqref{eqn:residue_orth}, 
we note that $|| \rv^\prime_a || = || \av ||,$ which is also the maximum value of $|| \rv^\prime_a ||$.
Therefore, such an MTD achieves the maximum possible value of $P^{\prime}_D (\av).$

\section*{Appendix C: Conjecture of Section~5.3 }
In this appendix, we present arguments that the attack detection probability $P^\prime_D (\av)$
increases as we select MTD perturbations with higher $\gamma(\Hm,\Hm^\prime).$
We use the short-hand notation $f(\uv,\vv)$ to represent the quantity $\dsp \max_{ \substack {\uv \in \mathcal{F},\uv \in \mathcal{G} \\ ||\uv|| = 1, ||\vv|| = 1}} |\uv^H \vv|.$
\vspace{2 mm}

The conjecture can be argued by examining the dependence of $||\rv^{\prime}_a||$ on $\gamma(\Hm,\Hm^{\prime})$ in the following three cases: 
\begin{itemize}
\item { Case~1:} When $Col(\Hm^\prime)$ is the orthogonal complement of $Col(\Hm),$
we have that $f (\uv, \vv) = 0$ (since $\uv^H \vv = 0, \ \forall \uv \in Col(\Hm), \vv \in Col(\Hm^{\prime})$), and $\gamma(\Hm,\Hm^{\prime}) = cos^{-1} (0) = \pi/2.$ From the arguments in Appendix~B, recall that in this case, $||\rv^\prime_a|| = ||\av||.$ 

\item { Case~2:} When $Col(\Hm)$ and $Col(\Hm^\prime)$ are identical (e.g. when $\Hm^\prime = (1+\eta) \Hm$), we have that $f (\uv, \vv) = 1,$ and $\gamma(\Hm,\Hm^{\prime}) = cos^{-1} (1) = 0$.
In this case, after straightforward simplification, it can be shown that $||\rv^\prime_a|| = 0.$

\item { Case~3:} For $0 \leq \gamma \leq \pi/2,$ from reference \cite{TeixeiraCDC2010}, we have the following bound
\begin{align}
||\rv^{\prime}_a|| \leq \sin (\gamma(\Hm,\Hm^{\prime})) ||\av||. \label{eqn:res_attack_bd}
\end{align} 
Note that the bound of \eqref{eqn:res_attack_bd} increases as $\gamma(\Hm,\Hm^{\prime})$ increases, which 
suggests that $||\rv^{\prime}_a||$ also increases. 
\end{itemize}
The conjecture can be justified from the observation in these three cases and using the fact that $P^\prime_D (\av)$ increases as 
$||\rv^{\prime}_a||$ increases (Appendix~B).

\end{document}